\documentclass[11pt,a4paper]{article}
\pdfoutput=1
\usepackage{jheppub}
\usepackage{amsmath,epsfig}

\usepackage{mathrsfs}
\usepackage{amssymb}
\usepackage{mathtools}
\usepackage{graphics}
\usepackage{slashed}
\usepackage{url}
\usepackage{hyperref}
\usepackage{enumitem}
\usepackage{bbold}

\usepackage{color}
\usepackage[normalem]{ulem}

\usepackage{multirow}
\usepackage[svgnames,dvipsnames,x11names,table]{xcolor}
\usepackage[framemethod=default]{mdframed}
\newmdenv[skipabove=7pt,
skipbelow=7pt,
rightline=true,
leftline=true,
topline=true,
bottomline=true,
backgroundcolor=gray!7,
linecolor=gray,
innerleftmargin=5pt,
innerrightmargin=5pt,
innertopmargin=0pt,
innerbottommargin=10pt,
leftmargin=0cm,
rightmargin=0cm,
linewidth=1.5pt]{eBox}

\usepackage{orcidlink}
\newcommand{\orcid}[1]{\orcidlink{#1}}
\usepackage{xcolor}
\usepackage[T1]{fontenc}
\usepackage[table]{xcolor}
\definecolor{tabred}{rgb}{0.8392156862745098, 0.15294117647058825, 0.1568627450980392}

\usepackage{scalerel}
\newcommand{\pmns}{U_{\scaleto{\mathrm{PMNS}}{4pt}}}
\newcommand{\mnus}{m_{\nu_{\mathrm{s}}}}
\newcommand{\mnu}{m_{\nu_{\mathrm{lightest}}}}
\newcommand{\mnua}{m_{\nu_{\mathrm{a}}}}
\newcommand{\tnu}{T^{0\nu}_{1/2}}

\newcommand\barparen[1]{\overset{%
   \scriptscriptstyle(-)}{#1}}

\title{Sterile neutrino dark matter within the $\nu$SMEFT}

\author[a]{Kaori Fuyuto \orcid{0000-0002-6924-0665},}
\author[a]{Jacky Kumar \orcid{0000-0001-9053-0731},}
\author[a]{Emanuele Mereghetti \orcid{0000-0002-8623-5796},}
\author[a]{Stefan Sandner \orcid{0000-0002-1802-9018},}
\author[a]{and Chen Sun \orcid{0000-0002-5145-1536}}

\affiliation[a]{Theoretical Division, Los Alamos National Laboratory, Los Alamos, NM 87545, USA}

\emailAdd{kfuyuto@lanl.gov}
\emailAdd{jacky.kumar@lanl.gov}
\emailAdd{emereghetti@lanl.gov}
\emailAdd{stefan.sandner@lanl.gov}
\emailAdd{chensun@lanl.gov}

\abstract{
Sterile neutrinos with masses at the $\mathrm{keV}$ scale and mixing to the active neutrinos offer an elegant explanation of the observed dark matter (DM) density. 
However, the very same mixing inevitably leads to radiative photon emission and the non-observation of such peaked $X$-ray lines rules out this minimal sterile neutrino DM hypothesis. 
We show that in the context of the Standard Model effective field theory with sterile neutrinos ($\nu$SMEFT), higher dimensional operators can produce sterile neutrino DM in a broad range of parameter space. 
In particular, $\nu$SMEFT interactions can open the large mixing parameter space due to their destructive interference, through operator mixing or matching, in the $X$-ray emission.
We also find that, even in the zero mixing limit, the DM density can always be explained by $\nu$SMEFT operators.
The testability of the studied $\nu$SMEFT operators in searches for electric dipole moments, neutrinoless double beta decay, and pion decay measurements is discussed.
}

\keywords{Neutrino physics, Neutrino Standard Model Effective Field Theory, Dark Matter}

\begin{document}
\preprint{LA-UR-24-23866}

\maketitle


\section{Introduction}
\label{sec:introduction}
The observation of neutrino oscillations establishes $\gg5\sigma$ evidence for beyond the Standard Model (BSM) physics~\cite{Esteban:2020cvm, deSalas:2020pgw}.
Furthermore, astrophysical and cosmological observations demonstrate that $\sim 26\%$ of the energy density of our Universe is attributed to a BSM matter component~\cite{Planck:2018vyg}.
It is well known that the introduction of sterile neutrinos in the Standard Model (SM) can solve both problems simultaneously, while being minimal in the theoretical realization~\cite{Asaka:2005an, Asaka:2005pn, Dodelson:1993je}.
The light neutrino mass spectrum in the arguably most minimal framework, the type-$\mathrm{I}$ seesaw~\cite{Minkowski:1977sc, Yanagida:1979as, Gell-Mann:1979vob, Mohapatra:1979ia}, can be generated by two sterile neutrinos, which mix to the active sector via Yukawa interactions.
It was first pointed out by Dodelson and Widrow (DW) that the inclusion of a third sterile neutrino ($\nu_s$) with mass at the $\mathrm{keV}$ scale and small mixing can account for the DM density~\cite{Dodelson:1993je}.
Its non-thermal production in the minimal realization fixes a unique relation between the active-sterile neutrino mixing, $\theta$, and the sterile neutrino mass $\mnus$:
\begin{align}
\label{eq:DW_mixing_intro}
    \theta^2_{\mathrm{DW}} \simeq 2 \times 10^{-8} \left( \frac{\mathrm{keV}}{\mnus} \right)^2\,.
\end{align}
However, for the $\nu_s$ to be DM, the resulting mixing is experimentally excluded due to non-observations of $X$-rays arising from the radiative decay $\nu_s \to \nu_a \gamma$~\cite{Pal:1981rm, Ng:2019gch, Roach:2022lgo}.
A plethora of proposals extending the DW mechanism to circumvent the $X$-ray constraint exist in literature and they can be divided into the following two classes:
i) Lower the mixing of Eq.~\eqref{eq:DW_mixing_intro} required to meet the correct DM abundance. This can be achieved for example by introducing additional interactions of the sterile neutrino~\cite{DeGouvea:2019wpf, Kelly:2020pcy, Fuller:2024noz, Johns:2019cwc, Astros:2023xhe, Bringmann:2022aim, An:2023mkf, Koutroulis:2023fgp, Holst:2023hff, Jaramillo:2022mos, Benso:2021hhh, Cho:2021yxk} or via its resonant production in a background of large lepton asymmetries~\cite{Shi:1998km}.
ii) For mixings as large as those of Eq.~\eqref{eq:DW_mixing_intro} additional sterile neutrino interactions lead to a destructive interference on the amplitude level of $\nu_s \to \nu_a \gamma$~\cite{Benso:2019jog}.

Current analyses in the literature so far have been limited to specific models and assumptions. 
In this paper we propose an agnostic approach to the sterile neutrino DM phenomenology within the context of effective field theories (EFTs).
In particular, we consider the Standard Model Effective Field Theory (SMEFT) \cite{Buchmuller:1985jz,Grzadkowski:2010es}, extended by sterile neutrinos ($\nu$SMEFT) \cite{delAguila:2008ir,Bhattacharya:2015vja,Liao:2016qyd,Bischer:2019ttk},
and study the effects of sterile neutrino interactions including up to dimension-$6$ operators.
Both modifications of the original DW mechanism as pointed out above in i) and ii) are found in the $\nu$SMEFT context.
Special attention is paid to $\nu$SMEFT operators leading to neutrino photon dipoles at low energies.
By generalizing to dipole operators the $1$-loop renormalization group equations (RGEs)~\cite{Chala:2020pbn, Datta:2020ocb, Datta:2021akg} and $1$-loop matching corrections at the electroweak scale (EW)~\cite{Dekens:2019ept, Cirigliano:2021peb}, we find that seven out of sixteen of the dimension-6 $\nu$SMEFT operators can induce sizable neutrino photon dipoles.  
For each of them, we identify the new physics scale at which $X$-ray constraints are evaded while the sterile neutrino can be DM.

The paper is organized as follows. 
In Sec.~\ref{sec:eft} we first provide a concise review of neutrino masses and discuss the diagonalization of the neutrino mass matrix. Then we introduce the $\nu$SMEFT basis, derive the RGEs 
for the mixing of dimension-$6$ operators onto dipoles
in $\nu$SMEFT, calculate matching corrections at the EW scale, and discuss additional renormalization group evolution from the EW scale to low energy in the sterile neutrino extended Low Energy EFT ($\nu$LEFT).
In Sec.~\ref{sec:sterile_nu_dm} the sterile neutrino DM production via mixing and the associated constraints are discussed.
Then we derive the DM production via dimension-$6$ $\nu$SMEFT interactions and confront the found parameter space to $X$-ray constraints on sterile neutrino DM.
Special attention is paid to scenarios in which $X$-ray emission i) is suppressed due to destructive interference effects and ii) arises solely from tree-level $\nu$SMEFT interactions. 
For both cases we demonstrate that the sterile neutrinos can match the observed DM abundance.
Sec.~\ref{sec:lab} considers possible laboratory probes for the $\nu$SMEFT interactions that can lead to sterile neutrino DM.
In Sec.~\ref{sec:conclu} we conclude.


\section{Neutrino masses, $\nu$SMEFT and $\nu$LEFT}
\label{sec:eft}
In this section we discuss neutrino masses and $\nu$SMEFT renormalization group (RG) running and its matching onto $\nu$LEFT.
\subsection{Neutrino mass matrix and diagonalization} 
Extending the SM with $n\geq 2$ sterile neutrinos, $\nu_R$, which are singlets under the SM gauge groups, leads to the following renormalizable Lagrangian at dimension $4$:
\begin{align}
\label{eq:Lag4}
{\cal L}^{(\mathrm{d}=4)} = {\cal L}_{\rm SM} + i \bar{\nu}_{R}^i\slashed{\partial} \nu_{R}^i - \sum_{\alpha,i} \bar L^\alpha Y^{\alpha i}_{\nu} \tilde{H} \nu^i_R - \sum_{i,j=1}^n {1\over 2} \overline{\nu^c}^{i}_R M^{ij}_R \nu_R^j+ \mathrm{h.c.}\,.
\end{align}
$Y_{\nu}$ is a $3\times n$ complex matrix parametrizing the Yukawa interactions and $M_R$ is a $n \times n$ complex symmetric matrix giving rise to Majorana masses and breaks global lepton number.
The SM lepton doublet is denoted by $L$ and the conjugate of the SM Higgs doublet is $\tilde{H} = i \sigma_2 H^*$.
We denote the conjugate fermion fields as $\psi^c = C \bar{\psi}^T$ and $\overline{\psi^c} = \psi^T C$, with $C=i \gamma^0 \gamma^2$ the charge conjugation matrix.
In terms of UV-complete models involving sterile neutrino, for example, left-right symmetric models~\cite{Pati:1974yy, Mohapatra:1974hk, Senjanovic:1975rk} and leptoquark models~\cite{FileviezPerez:2013zmv, Dorsner:2016wpm}, it is natural to assume that other new particles that interact with $\nu_R$ are also present. 
If these particles are relatively heavy, which is motivated by current collider experiments, such interactions are expressed by higher-dimensional, gauge-invariant non-renormalizable operators at low energy:
\begin{align}\label{eq:nuSMEFT}
    \mathcal{L}^{\nu\mathrm{SMEFT}} = \mathcal{L}^{(\rm d=4)} + \sum_{\mathrm{d}=5} \sum_{I} \frac{\mathcal{C}_I^{\mathrm{(d)}}}{\Lambda^{\mathrm{d}-4}} {\mathcal{O}_I}^{\mathrm{(d)}}\,,
\end{align} 
where operators $\mathcal{O}^{(\mathrm{d}\geq 5)}$ are constructed from the SM fields plus $\nu_R$, $\mathcal{C}_I^{(\mathrm{d})}$ denote the dimensionless Wilson coefficients of mass dimension $\mathrm{d}$ weighted by the EFT cut-off scale $\Lambda$. 
As we will see in the following, some interactions in $\mathcal{L}^{\nu\mathrm{SMEFT}}$ can generate light neutrino masses.

After EW symmetry breaking, the full neutrino mass matrix can be written as
\begin{align}
\label{eq:Mnu_def}
{\cal L}_{\rm mass}=-\frac{1}{2}\overline{N^c}M_{\nu}N+{\rm h.c.}\,,\hspace{1cm}
M_{\nu}=
\begin{pmatrix}
m_L & m_D^* \\
m_D^{\dagger} & M_R
\end{pmatrix}
\,,
\end{align}
with $N=(\nu_L, \nu_R^c)^T$. Here $m_D = v Y_\nu/\sqrt{2}$, while $m_L$ is induced by the higher dimensional operators in Eq. \eqref{eq:nuSMEFT}, as will be shown explicitly in Eq.~\eqref{eq:mL_induced}.
We can always choose a basis in which $M_R$ is real and diagonal.
To find the mass eigenstates of the neutrino tuple $N$, the matrix $M_\nu$ has to be diagonalized by a unitary matrix $U$, i.e.
\begin{align}
U^TM_{\nu}U={\rm diag}(m, M)\,,
\end{align}
where $m$ $(M)$ denote the active (sterile) neutrino mass eigenstates.
The unitary matrix is well approximated by~\cite{Blennow:2011vn}
\begin{align}
\label{eq:unitary_mixing}
    U \simeq \begin{pmatrix}
    \pmns & \theta \\
    - \theta^\dagger \pmns & 1
    \end{pmatrix}
    \,,
\end{align}
which assumes that the mixing in the active sector is nearly unitary (which is justified at the (sub-) percent level precision~\cite{Antusch:2006vwa, Fernandez-Martinez:2007iaa, Blennow:2023mqx}) and described by the Pontecorvo-Maki-Nakagawa-Sakata (PMNS) matrix $\pmns$,\footnote{We follow the parametrization for $\pmns$ as given in Chapter 14 \textit{Neutrino Masses, Mixing, and Oscillations} of the particle data group~\cite{ParticleDataGroup:2022pth}.} as well as the mixing between sterile and left-handed neutrinos is small and characterized by the $3\times n$ matrix $\theta$.
The relation to the physical neutrino masses thus induces the following relation~\cite{Huang:2013kma, Donini:2012tt}
\begin{align}
\label{eq:mL_diagonalization}
    \pmns^{\phantom{T}} m \pmns^T &= m_L^\dagger - \theta M_R \theta^T\,,\\
\label{eq:mD_diagonalization}
    \theta M - m_L^\dagger \theta^* &= m_D \,,\\
\label{eq:mR_diagonalization}
    M &= M_R\,.
\end{align}
Further assuming that $m_L^\dagger \ll \theta M_R \theta^T$,\footnote{This is justified at tree level if the dimension-$5$ Weinberg operator is located at a high scale and at loop level in the context of the present analysis as we will show in Eq.~\eqref{eq:mL_induced} together with the numerical result of Sec.~\ref{subsubsec:xray_cancellation}.}
we can express the active-sterile neutrino mixing as
\begin{align}
\label{eq:CI_mixing}
    \theta = i \pmns \sqrt{m} R^\dagger M^{-1/2}\,.
\end{align}
Here $R$ denotes a $n \times 3$ complex orthogonal matrix and can be generically parametrized via a rotation matrix with $(n^2-n)/2$ complex angles.
For the case of $n=3$ the explicit form of $R$ is given in Eq.~\eqref{app:eq:r_matrix}.
This allows to express the mixing of the active to sterile neutrinos via a parametrization which directly accounts for the constraints from neutrino oscillation experiments.
A derivation of this so known Casas-Ibarra parametrization~\cite{Casas:2001sr} is given in App.~\ref{app:casas_ibarra}.

Within the scope of the present paper, we will focus on the explicit scenario with $n=3$ sterile neutrinos and assume $\mnus \equiv m_{\nu_4} \ll m_{\nu_5} \leq m_{\nu_6}$.
In particular, the two heavier sterile neutrinos induce two masses for the light neutrinos and therefore explain the measured light neutrino mass splitting~\cite{Esteban:2020cvm, deSalas:2020pgw}.
On the other hand, the lightest sterile neutrino is assumed to have mass at the $\mathcal{O}(\mathrm{keV})$ scale and thus can be a DM candidate, as will be shown in Sec.~\ref{sec:sterile_nu_dm}.
This lightest sterile neutrino represents the main focus of the analysis.
Note that we can assume without loss of generality the sterile neutrino $\nu_4$ to be the DM candidate for normal hierarchy (NH) and inverted hierarchy (IH). 
For the IH scenario, in the mixing of Eq.~\eqref{eq:CI_mixing} we thus need to replace $R^\dagger \mapsto R^\dagger P$ with a $123 \mapsto 312$ permutation matrix
\begin{align}
    P = \begin{pmatrix}
    0 & 1 & 0 \\
    0 & 0 & 1 \\
    1 & 0 & 0
    \end{pmatrix}
    \,.
\end{align}

\subsection{$\nu$SMEFT: operator basis and running}
The renormalizable Lagrangian in Eq.~\eqref{eq:Lag4} can be extended to include the most general set of interactions invariant under the SM 
$SU(3)_c \times SU(2)_L \times U(1)_Y$ gauge group and the Lorentz group, organized according to their canonical dimension, which leads to the Lagrangian of Eq.~\eqref{eq:nuSMEFT}.
The operators of lowest dimension arise at dimension $5$, and are suppressed by one power of new physics scale $\Lambda$. Odd-dimension operators in $\nu$SMEFT break lepton number~\cite{Helset:2019eyc}, in particular,
dimension-$5$ operators break lepton number by two units. 
Without sterile neutrinos, one can write the Weinberg operator \cite{Weinberg:1979sa}, which at low-energy gives rise to Majorana masses for left-handed neutrinos. 
In addition, it is possible to write a transition magnetic moment for sterile neutrinos and a dimension-$5$ correction to the the sterile neutrino Majorana mass 
\begin{align}
\label{eq:Lag_dim5}
    \mathcal L^{(5)} = \frac{C_5}{\Lambda} \varepsilon_{k l} \varepsilon_{mn} \overline{L^c}_k^T   L_m H_l H_n  + \frac{C^R_{\nu B}}{\Lambda}  \overline{\nu^c}_R \, \sigma^{\mu \nu} \nu_R B_{\mu \nu} + \frac{C_{\nu H}}{\Lambda}
    \overline{\nu^c}_R \nu_R H^\dagger H\,,
\end{align}
where the coefficients carry flavor indices, which we do not indicate.
\begin{table}
\centering
\resizebox{\columnwidth}{!}{
\begin{tabular}{|c|c|c|c|c|c|}
\hline
 \multicolumn{2}{|c|}{$\psi^2H^3$} & \multicolumn{2}{|c|}{$\psi^2H^2D$} & \multicolumn{2}{|c|}{$\psi^2HX(+\mbox{H.c.})$}
\\
\hline
$\mathcal{O}_{L\nu H}(+\mbox{H.c.})$ & $(\bar{L}\nu_R )\tilde{H}(H^\dagger H)$ &
$\mathcal{O}_{H\nu }$ &  $(\bar \nu_R \gamma^\mu \nu_R )(H^\dagger i \overleftrightarrow{D_\mu} H)$&
$\mathbf{\color{tabred}{^*\, \mathcal{O}_{\nu B} }}$ & $(\bar{L}\sigma_{\mu\nu}\nu_R )\tilde{H}B^{\mu\nu}$
\\
& &
$\mathbf{\color{tabred}{^*\, \mathcal{O}_{H\nu e}}}$ & $(\bar{\nu}_R \gamma^\mu e)({\tilde{H}}^\dagger i D_\mu H)$ &

$\mathbf{\color{tabred}{^*\, \mathcal{O}_{\nu W}}}$ &$(\bar{L}\sigma_{\mu\nu}\nu_R )\tau^I\tilde{H}W^{I\mu\nu}$
\\
\hline
\multicolumn{2}{|c|}{$(\bar{R}R)(\bar{R}R)$}  &   \multicolumn{2}{|c|}{$(\bar{L}L)(\bar{R}R)$} &   \multicolumn{2}{|c|}{$(\bar{L}R)(\bar{L}R)$ and $(\bar{L}R)(\bar{R}L)$ $(+\mbox{H.c.})$}
\\
\hline
$\mathcal{O}_{\nu \nu }$ & $(\bar \nu_R \gamma^\mu \nu_R )(\bar \nu_R \gamma_\mu \nu_R )$ &
$\mathcal{O}_{L\nu }$ & $(\bar{L}\gamma^\mu L)(\bar \nu_R \gamma_\mu \nu_R )$ &
$ \mathbf{\color{tabred}{^*\,\mathcal{O}_{L\nu Le}}} $ & $(\bar{L}^i\nu_R )\epsilon^{ij}(\bar{L}^je)$
\\
$\mathcal{O}_{e\nu }$ & $(\bar{e}\gamma^\mu e)(\bar \nu_R \gamma_\mu \nu_R )$ &
$\mathcal{O}_{Q\nu }$ & $(\bar{Q}\gamma^\mu Q)(\bar \nu_R \gamma_\mu \nu_R )$ &
$\mathbf{\color{tabred}{^*\, \mathcal{O}_{L\nu Qd}^{(1)}}}$ & $(\bar{L}^i\nu_R )\epsilon^{ij}(\bar{Q}^jd)$
\\
$\mathcal{O}_{u\nu }$ & $(\bar{u}\gamma^\mu u)(\bar \nu_R \gamma_\mu \nu_R )$ &
& &
$\mathbf{\color{tabred}{^*\, \mathcal{O}_{L\nu Qd}^{(3)}}}$ & $(\bar{L}^i\sigma^{\mu\nu}\nu_R)\epsilon^{ij}(\bar{Q}^j\sigma_{\mu\nu}d )$
\\
$\mathcal{O}_{d\nu }$ & $(\bar{d}\gamma^\mu d)(\bar \nu_R \gamma_\mu \nu_R )$&
& &
&
\\
$\mathbf{\color{tabred}{^*\, \mathcal{O}_{du\nu e}}}$ & $ (\bar{d}\gamma^\mu u)(\bar \nu_R \gamma_\mu e)$&
& &
$\mathcal{O}_{Qu\nu L}$ &$(\bar{Q}^iu)(\bar \nu_R L^i)$\\
\hline
\end{tabular}
} 
\caption{The dimension-six $\nu$SMEFT operators involving $\nu_R$ (with $L$ and $B$ conserved)~\cite{Bischer:2019ttk,Liao:2016qyd,Datta:2021akg}.  
The operators highlighted in bold, red and marked with a star ($^*$) in front can lead to a relevant photon dipole operator.
}
\label{tab:nuSMEFT_operators_dim6}
\end{table}

The complete set of dimension-$6$ operators with sterile neutrinos was constructed in Refs.~\cite{Liao:2016qyd,Datta:2021akg}, and we summarize them in Tab.~\ref{tab:nuSMEFT_operators_dim6}.
At tree level, the operators in Tab.~\ref{tab:nuSMEFT_operators_dim6}
can influence many processes, from $\beta$ decay and neutrinoless double beta decay ($0\nu\beta\beta$),
to neutrino-nucleus scattering, to $Z$, $W$ and Higgs decays.
Here we are mostly interested in operators that can contribute to $X$-ray spectra of galaxies and galaxy clusters, in the case in which the lightest sterile neutrino constitutes DM.
These corrections 
are dominantly induced by the low-energy Dirac photon dipole operator, 
\begin{equation}
   \mathcal L_{\rm dipole} =  \frac{1}{v} 
   C_{\substack{\nu F \\r s}} ~\bar{\nu}^r_{L} \sigma^{\mu\nu}\nu_R^s F_{\mu\nu} + \mathrm{h.c.}\,.
\end{equation}
As we will see in Sec.~\ref{subsubsec:xray_cancellation}, non-observation of $X$-rays gives an extremely strong bound on the above photon dipole operator, e.g., $|C_{\nu F}|< {\cal O}(10^{-17})$ for $m_{\nu_s}=100\,\mathrm{keV}$.
This corresponds to $\Lambda > {\cal O}(10^{15})\,\mathrm{GeV}$ assuming $C_{\nu F}/v=1/\Lambda$. 
In what follows, we discuss all possible  contributions from dimension-$6$ $\nu$SMEFT operators listed in Tab.~\ref{tab:nuSMEFT_operators_dim6} to the neutrino dipole operators employing full $1$-loop RG running and $1$-loop matching.\footnote{Our analysis corresponds to next-to-leading-log (NLL)~\cite{Buchalla:1995vs, Kumar:2024yuu} order in RGE and matching. Leading-log (LL) contributions arise from the solution of the $1$-loop RGEs in $\nu$SMEFT and its low-energy version, $\nu$LEFT, while NLL corrections are induced by $1$-loop matching corrections at the EW scale. To reach full NLL accuracy, knowledge of the $2$-loop anomalous dimension is required. 
These corrections are beyond the scope of this work.}

The photon dipole operator receives a tree level contribution from the operators $\mathcal O_{\nu W}$ and $\mathcal O_{\nu B}$ after EW symmetry breaking. 
In the flavor basis 
\begin{equation}
C_{\substack{\nu F \\ rs}} = \frac{v^2}{\sqrt{2}\Lambda^2}\left(s_{w}C_{\substack{\nu W\\ rs}}+c_w C_{\substack{\nu B\\ rs}}\right),
\end{equation}
where $s_w = \sin \theta_w$ and $c_w = \cos \theta_w$, with $\theta_w = 0.492262\,\mathrm{rad}$ the Weinberg angle~\cite{ParticleDataGroup:2022pth}.
The linear combination
$c_w C_{\nu W} - s_w C_{\nu B}$
induces dipole couplings to the $Z$ and $W$ bosons, but, at least at tree level, not to the photon.

Many more contributions can be induced via renormalization group effects.
Most of the anomalous dimension matrix was derived in Refs.~\cite{Chala:2020pbn, Datta:2020ocb, Datta:2021akg}.
In addition to the contributions discussed in these papers, here we need the mixing of four-fermion onto dipole operators,
which is proportional to the product of gauge and Yukawa couplings.
This mixing can be obtained from the results of Ref.~\cite{Jenkins:2013wua}, by replacing $L \leftrightarrow Q$, $d \leftrightarrow e$, $u \leftrightarrow \nu_R$ along with simultaneous adjustment of the couplings. 
Neglecting neutrino Yukawa interactions, we obtain the RGEs
\begin{align}
 \label{eq:nuW}
 \begin{split}
 \dot C_{\substack{\nu W \\rs}}&= \left[ (3 c_{F,2} - b_{0,2}) g_2^2  - 3 {\rm y}_\ell^2  g_1^2  + \mathrm{Tr}^2  + {5\over 2} Y_e^2   \right]  C_{\substack{\nu W \\rs}}  + 3 g_1 g_2 {\rm y}_{\ell} C_{\substack{\nu B \\rs}}  \\&
 + \frac{g_2}{4}C_{\substack{L\nu Le \\ psrt}}\left[Y_e^{\dagger} \right]_{tp} 
 + 2g_2N_CC^{(3)}_{\substack{L\nu Qd \\rspt}}\left[Y_d^{\dagger} \right]_{pt}\,,
 \end{split}\\
\label{eq:nuB}
\begin{split}
\dot C_{\substack{\nu B \\rs}}&=
\left[- 3 c_{F,2} g_2^2 + ( 3 {\rm y}_\ell^2 - b_{0,1}) g_1^2 + \mathrm{Tr}^2 -{3\over 2 } Y_e^2 \right] C_{\substack{\nu B \\rs}} + 12 c_{F,2} g_1 g_2 {\rm y}_{\ell} \, C_{\substack{\nu W \\rs}} \\
&-\frac{g_1}{2}({\rm y}_e +{\rm y}_{\ell})C_{\substack{L\nu Le \\psrt}}\left[Y_e^{\dagger} \right]_{tp} -4g_1N_C({\rm y}_d+{\rm y}_Q)C^{(3)}_{\substack{L\nu Qd \\rspt}}\left[Y_d^{\dagger} \right]_{pt}\,,
\end{split}\\
\label{eq:LnuLe}
\dot C_{\substack{L\nu Le \\rspt}}&= 
\left[ ({\rm y}_e^2 - 8 {\rm y}_e {\rm y}_l + 6 {\rm y}_{\ell}^2) g_1^2 - \frac{3}{2} g_2^2 \right ]C_{\substack{L\nu Le \\rspt}}
-  (4 {\rm y}_{\ell} ({\rm y}_e + {\rm y}_{\ell}) g_1^2 - 3 g_2^2) C_{\substack{L\nu Le \\psrt}}
\,,\\
\label{eq:LnuQd_tensor}
\begin{split}
\dot C^{(3)}_{\substack{L\nu Qd \\rspt}}&= \left(2 \left({\rm y}_d^2 - {\rm y}_d {\rm y}_\ell - 2 {\rm y}^2_{\ell}\right) g_1^2  - 3 g_2^2 + 2 c_{F,3} g_3^2 \right) C^{(3)}_{\substack{L\nu Qd \\rspt}} \\
&+ \frac{1}{8} \left( - 4 {\rm y}_{\ell} (2 {\rm y}_d - {\rm y}_{\ell}) g_1^2 + 3 g_2^2 \right) C^{(1)}_{\substack{L\nu Qd \\rspt}} -\frac{1}{2}\left[Y_e \right]_{rw}\left[Y_u \right]_{pv}C^{*}_{\substack{du \nu e \\tvsw}} \,,
\end{split}\\
\label{eq:LnuQd_scalar}
\begin{split}
\dot C^{(1)}_{\substack{L\nu Qd \\rspt}}&=
-\left( 6 \left({\rm y}_d^2 - {\rm y}_d {\rm y}_\ell\right) g_1^2  - 3 g_2^2 + 6 c_{F,3} g_3^2 \right) C^{(1)}_{\substack{L\nu Qd \\rspt}} \\
&- \left(  24 {\rm y}_{\ell} (2 {\rm y}_d - {\rm y}_{\ell}) g_1^2 - 18 g_2^2 \right) C^{(3)}_{\substack{L\nu Qd \\rspt}} \,,
\end{split} \\
\label{eq:LnuuQ}
\begin{split}
\dot C_{\substack{ d u \nu e \\ stpr}} &= [Y_e]_{vr} [Y_d^{\dagger}]_{sw} ~ C_{\substack{Qu\nu L \\ wtpv}}\,,
\end{split}
\end{align}
where $\dot{C} = (16 \pi^2)~d C/ d \ln \mu$. 
The  mixing of $C_{\nu W}$, $C_{\nu B}$ with four-fermion operators in Eqs. \eqref{eq:nuW}
and \eqref{eq:nuB} is a new result of this work,  while the rest of RGEs are taken from Refs. \cite{Chala:2020pbn, Datta:2020ocb, Datta:2021akg}.
The  $SU(2)_L$ and $SU(3)_c$ Casimir are $c_{F,2} = 3/4$ and $c_{F,3} = (N_C^2 - 1)/2 N_C$, while the $U(1)_Y$ and $SU(2)_L$ $\beta$ functions
\begin{align}
b_{0,1} = - \frac{1}{6} - \frac{20}{9} n_g\,, \quad b_{0,2}  = \frac{43}{6} - \frac{4}{3} n_g\,,
\end{align}
with $n_g=3$ the number of fermion generations.
${\rm y}_f$ denotes the hypercharge
\begin{align}
    {\rm y}_q = \frac{1}{6}\,, \quad
    {\rm y}_u = \frac{2}{3}\,, \quad {\rm y}_d = -\frac{1}{3}, \quad 
    {\rm y}_{\ell} = - \frac{1}{2}, \quad
    {\rm y}_e = -1\,,
\end{align}
and $\mathrm{Tr}^2 = 3 \mathrm{Tr}(Y_u^2 + Y_d^2) + \mathrm{Tr}(Y_e^2)$.
In the solution of the RGEs, we include the QCD running of the quark Yukawa couplings and of the strong coupling $g_3$
\begin{align}
\label{eq:y_g_running}
    \dot{Y}_q = - 6 c_{F,3} g_3^2 Y_q\,, \qquad \dot{g}_3  = -  b_{0,3} g_3^3\,, \qquad  b_{0,3} = 11 - \frac{2}{3} n_f\,,
\end{align}
with $n_f = 6$ above the top mass.
We neglect the running of the lepton Yukawas and of the EW couplings.
The initial conditions for Eq.~\eqref{eq:y_g_running} are the quark masses from~\cite{ParticleDataGroup:2022pth} and $\alpha_s(m_Z^2) = 0.1184$~\cite{ParticleDataGroup:2022pth}.

\begin{figure}[!t]
\centering
\includegraphics[width=0.975 \textwidth]{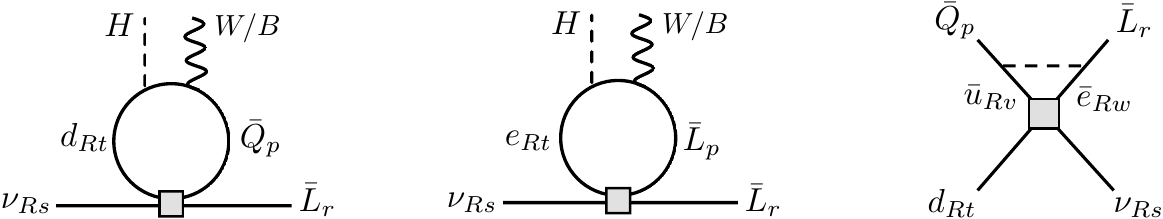}
\caption{Examples of operator mixing. A gray square corresponds to an insertion of a $\nu\mathrm{SMEFT}$ operator. The left and middle diagrams represent mixing from ${\cal O}_{L\nu Qd}$ and ${\cal O}_{L\nu Le}$, and the right diagram depicts the generation of ${\cal O}_{L\nu Qd}$ from ${\cal O}_{du\nu e}$. Lower indices correspond to flavor.  }
\vspace{-1em}
\label{fig:operator_mixing}
\end{figure}

In Eq.~\eqref{eq:LnuQd_tensor} we are neglecting contributions proportional to the neutrino Yukawa couplings, which induce very small correction to the dipole operators. 
For $C_{du\nu e}$, we only include the mixing 
with  $C_{Qu\nu L}$,
and neglect contributions proportional to the operators already included in Eqs.~\eqref{eq:nuW}-\eqref{eq:LnuQd_scalar}.
With these assumptions Eqs.~\eqref{eq:nuW}-\eqref{eq:LnuuQ} 
provide the LL contribution to $C_{\nu F}$. 
In the limit of $Y_\nu \mapsto 0$, no other operators in Tab.~\ref{tab:nuSMEFT_operators_dim6} feeds into $C_{\nu B}$ or $C_{\nu W}$ at $1$-loop, and in this sense the set of Eqs.~\eqref{eq:nuW}-\eqref{eq:LnuuQ} is closed.
Some examples of operator mixing in terms of Feynman diagrams are shown in Fig.~\ref{fig:operator_mixing}. 
The left and middle diagrams show mixing from ${\cal O}^{(3)}_{L\nu Qd}$ and ${\cal O}_{L\nu Le}$ to ${\cal O}_{\nu W/\nu B}$ described in Eq.~\eqref{eq:nuW} and \eqref{eq:nuB}. 
The operator ${\cal O}_{du\nu e}$ first induces ${\cal O}^{(3)}_{L\nu Qd}$ as in Eq.~\eqref{eq:LnuQd_tensor}, leading to
dipole operators in the end. The right diagram in Fig.~\ref{fig:operator_mixing} corresponds to the first step through two Yukawa interactions.

As we will see in Sec.~\ref{subsec:xray}, $X$-ray constraints are so strong that one might actually wonder about contribution induced by the neutrino Yukawa coupling $Y_\nu$, and if neglecting these contributions in Eqs.~\eqref{eq:nuW}-\eqref{eq:LnuuQ} is justified.
Considering only operators that run into the purely leptonic or semileptonic tensor operators, we find
\begin{align}
\label{eq:LnuLeY}
\dot C_{\substack{L\nu Le \\rspt}}&= -4 \left(\left[Y^\dagger_{\nu} 
\right]_{rv} \left[Y^\dagger_{e} 
\right]_{pw} -\left[Y^\dagger_{\nu} 
\right]_{pv} \left[Y^\dagger_{e} 
\right]_{rw} \right)  C_{\substack{e\nu \\wt vs}}  + 4 \left[Y^\dagger_{\nu} 
\right]_{pw} \left[Y^\dagger_{e} 
\right]_{vt} C_{\substack{L \nu\\ r v w s}}\,,
 \\
\label{eq:LnuQd_tensorY}
\begin{split}
\dot C^{(3)}_{\substack{L\nu Qd \\rspt}}&= + \frac{1}{2}\left[Y_{\nu}\right]_{rw}\left[Y_d \right]_{vt}C^{*}_{\substack{Q\nu \\pvws}} +\frac{1}{2}\left[Y_{\nu} \right]_{rw}\left[Y_{d} \right]_{pw}C_{\substack{d \nu \\ wsvt}}\,.
\end{split}
\end{align}
The operators $C_{d \nu}$ and $C_{H \nu}$ feed into $C_{e\nu}$, $C_{L\nu}$, $C_{Q\nu}$ and $C_{d\nu}$ at one loop~\cite{Datta:2020ocb}, so that actually all operators in Tab.~\ref{tab:nuSMEFT_operators_dim6} generate a photon neutrino dipole at NLL, with the exception of $C_{\nu \nu}$
and $C_{L\nu H}$.\footnote{$C_{L\nu H}$ can generate a $Z$ dipole operator at two loops, via Barr-Zee diagrams \cite{Barr:1990vd,Barr:1990um}, which can then run into a photon dipole via Eqs.~\eqref{eq:nuW} and~\eqref{eq:nuB}.} 
The neutrino Yukawa path opened by Eqs.~\eqref{eq:LnuLeY} 
and~\eqref{eq:LnuQd_tensorY}, however, leads to constrains $\Lambda \lesssim v$, and is thus outside the regime of the validity of $\nu$SMEFT. 

In Sec.~\ref{sec:matching} we will show that the operator $C_{H\nu e}$ generates a matching correction to the 
dipole operator. It might therefore be important to consider its evolution from the scale of new physics to the EW scale, as well as mixing with other operators. It turns out that the $C_{H\nu e}$ operator exhibits only the self running or mixing with neutrino dipole operators at $1$-loop level. These are governed by the equation
\begin{align}
\label{eq:CHnue_rge}
\dot C_{H\nu e } & = - 3 g_1 Y_e C_{\nu B} + 9 g_2 Y_e C_{\nu W} + ( -3 g_1^2  + Y_e^2+ 2\textrm{Tr}^2) C_{H \nu e} \,,
\end{align}

In addition to the running on $\nu$SMEFT operators into the dipole, 
another interesting effect is that the
dipole operators runs into the dimension-5 Weinberg operator,
thus inducing a left-handed Majorana mass $m_L$.
The RGE is given by
\begin{align}
    \dot C_{\substack{5 \\p r}}  = \frac{12}{\Lambda}  \left(  g_1 {\rm y}_\ell C_{\substack{\nu B\\ s_1 p} }^{\dagger} + \frac{g_2}{2} C_{\substack{\nu W\\ s_1 p} }^{\dagger}\right) \left[M_R\right]_{s_1 s_2} \left[ Y_\nu^{\dagger} \right]_{s_2 r}\,.
\end{align}
Note, that in terms of physical fields, only the loop with a $Z$ boson exchange contributes.
Thus, the left-handed Majorana mass $m_L$ is only generated \textit{above} the EW phase transition and then remains constant.
Comparing to Eq.~\eqref{eq:Mnu_def} yields 
\begin{align}
\label{eq:mL_induced}
    m_L \simeq \frac{12}{16\pi^2} \frac{v}{\sqrt{2}} \frac{M^2}{\Lambda^2} \theta  \left(  g_1 {\rm y}_\ell C_{\substack{\nu B} }^{\dagger} + \frac{g_2}{2} C_{\substack{\nu W} }^{\dagger}\right) \log\left(\frac{\mu}{\Lambda}\right)\,,
\end{align}
with $\mu \sim m_t$.
We will see that, for operators that contribute to $X$-ray, this effect is too small to matter.
The dipoles also induce a correction to the Dirac mass, with the operator $C_{L\nu H}$ and a correction to the $Z \bar{\nu}_R \nu_R$ coupling, via $C_{H\nu}$.  These were derived in Ref. \cite{Chala:2020pbn, Datta:2020ocb, Datta:2021akg}, and we add them here for completeness
\begin{align}
\begin{split}
\dot C_{L \nu H } & = - 3 g_1 (g_1^2 + g_2^2) C_{\nu B} + 3 g_2 (g_1^2 + 3 g_2^2 + 4 Y_e^2) C_{\nu W} \\
&+ Y_e(3 g_2^2  - 2 Y_e^2) C_{H\nu e}+  \left (-{ 9 \over 4} g_1^2 - {27\over 4} g_2^2  - {3\over 2} Y_e^2  +3\textrm{Tr}^2 \right ) C_{ 
 L\nu  H} \,,
\end{split}\\
\begin{split}
\dot C_{H\nu } & =  3 g_1 Y_e C_{\nu B} - 9 g_2 Y_e C_{\nu W} + \left ({1 \over 3} g_1^2 + 2 \textrm{Tr}^2  \right ) C_{H \nu} + 2 Y_e^2 C_{H \nu e}.
\end{split}
\end{align}
We find that the generated operators on the l.h.s.  do not given any important constraints on the neutrino dipoles beyond the stringent $X$-rays.

\subsection{$\nu$LEFT: Matching and running}
\label{sec:matching}
\begin{figure}[!t]
\centering
\includegraphics[width=5cm]{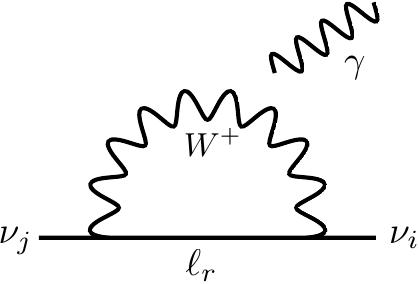}
\qquad 
\includegraphics[width=5cm]{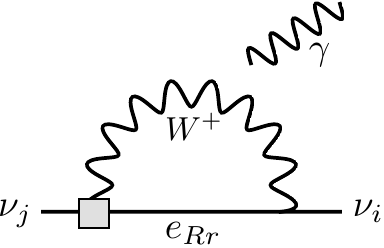}
\caption{
Matching contributions to the dipole operator in the $\nu$SM (left) and in the presence of dimension-six $\nu$SMEFT operators (right)
}
\vspace{-1em}
\label{fig:Xray_standard}
\end{figure}

At the EW scale, we integrate out the $W$, $Z$ and Higgs bosons, and match onto $\nu$LEFT. 
The $\nu$LEFT operators at dimension $5$ and $6$ have been derived in~\cite{Bischer:2019ttk,Chala:2020vqp}.
Here, we focus on the dipole operator, a leptonic scalar operator and a tensor semileptonic operator
\begin{equation}
    \mathcal L_{\rm LEFT} \supset  \frac{C_{\nu F}}{v} \bar \nu_L \sigma^{\mu \nu} \nu_R F_{\mu \nu} + \frac{{C}^{S,RR}_{e \nu \nu e}}{\Lambda^2}  \, \bar{\nu}_L  e_R \bar e_L \nu_R 
    + \frac{ {C}^{T,RR}_{\nu d } }{\Lambda^2} \,\bar{\nu}_L \sigma_{\mu \nu} \nu_R \, \bar d_L \sigma^{\mu \nu} d_R\,.
\end{equation}
After EW symmetry breaking,
\begin{align}
\label{eq:runEW}
    C_{\substack{\nu F\\ rs}}(\mu = m_t) =  \frac{v^2}{\sqrt{2}\Lambda^2}\left(s_{w}C_{\substack{\nu W\\ rs}}+c_w C_{\substack{\nu B\\ rs}}\right)(\mu = m_t)\,,
\end{align}
where the dipole operators on the l.h.s are given by the solution of the $\nu$SMEFT RGEs 
discussed in the previous section. 
For operators that contribute at LL,  we do not consider finite matching corrections, which contribute at NLL, and are thus subleading. 
We consider two important matching corrections. 
The first is the one induced by dimension-$4$ interactions,
through e.g. the left diagram of Fig.~\ref{fig:Xray_standard}.
This diagram gives~\cite{Pal:1981rm}
\begin{align}
\label{eq:sm}
C_{\substack {\nu F\\ rs}}=  \frac{e }{16 v \pi^2}\sum_{r=e,\mu,\tau} F(x_r) \left[m_D\right]_{r s}\,,
\end{align}
where the Dirac mass $m_D$ is defined in Eq.~\eqref{eq:mD_diagonalization}, $x_r={m_r^2}/{m^2_W}$ and the loop function $F$ is
\begin{align}
\label{eq:Fxr}
F(x)=\frac{1}{(1-x)^2}\Bigg[-\frac{3}{4}\left(2-5x+x^2 \right)+\frac{3}{2}\frac{x^2\ln x}{1-x} \Bigg]\,,
\end{align}
with $F(x_r)\simeq -3/2$ for $x_r\ll 1$.
In $\nu$SMEFT, $\nu_R$ couples to the $W$ boson and a right-handed electron via the operator $C_{H\nu e}$. This operator generates a dipole via the right diagram in Fig.~\ref{fig:Xray_standard}, which does not depend on the active-sterile mixing, and is proportional to the charged lepton mass, rather than the neutrino mass. We find~\cite{Cirigliano:2021peb}
\begin{align}
\label{eq:RH}
C_{\substack{\nu F\\ rs}}=-\frac{e}{2(4\pi)^2}\frac{v}{\Lambda^2}\sum_{r=e,\mu,\tau}  C^*_{\substack{H\nu e\\ sr}} w(x_r)m_r\,,
\end{align}
with
\begin{align}
w(x)=\frac{4+x(x-11)}{2(x-1)^2}+3\frac{x^2\ln x}{(x-1)^3}\,.
\end{align}
Using the results of SMEFT matching on LEFT~\cite{Dekens:2019ept},
and using the same mapping between SMEFT and $\nu$SMEFT operators discussed for the RGE, we found that all other matching contributions are suppressed by small neutrino Yukawa.
Eq.~\eqref{eq:runEW}, \eqref{eq:sm} and \eqref{eq:RH} are the initial condition for further evolution from the EW to scales relevant for $X$-ray emission.
Below the EW scale, the RGEs for the photon dipole operator are
\begin{align}
    \dot{C}_{\substack{\nu F \\ r s}} &= \frac{v}{\Lambda^2} e Q_e \left[M_e\right]_{w v} C^{S, RR}_{\substack{e \nu \nu e \\w s r v}} - \frac{v}{\Lambda^2} 8 e Q_d N_C\left[M_d\right]_{w v} C^{T,RR}_{\substack{\nu d \\r s v w}}\,,\\ 
   \dot{C}^{S,RR}_{\substack{e \nu \nu e \\r s p t}} &= 2  e^2 Q_e^2  {C}^{S,RR}_{\substack{e \nu \nu e \\r s p t}} \,,\\
   \dot{C}^{T,RR}_{\substack{\nu d \\r s p t}} &= 2 \left(e^2 Q_d^2  + g_3^2 c_{F,3}\right) {C}^{T,RR}_{\substack{\nu d \\r s p t}} \,,
\end{align}
with $Q_e = -1$ and $Q_d = -1/3$.
The tree level matching of the $\nu$SMEFT operators onto the $\nu$LEFT basis yields
\begin{align}
    C^{S, RR}_{e\nu \nu e} = - C^{}_{L\nu L e}\,, \quad
    C^{T, RR}_{\nu d} = C^{(3)}_{L\nu Q d} \,.
\end{align}

\begin{table}[!t]
\centering
\begin{tabular}{|c||c|c||c|c||c|c|}
\hline 
&   
$[C^{(3)}_{L\nu Q d}]_{11}$ & $[C^{(1)}_{L\nu Q d}]_{11}$ 
 &
$[C^{(3)}_{L\nu Q d}]_{22}$
 &
$[C^{(1)}_{L\nu Q d}]_{22}$ 
& 
$[C^{(3)}_{L\nu Q d}]_{33}$
&
$[C^{(1)}_{L\nu Q d}]_{33}$
\\
\hline
$  [C_{\nu F}]_{\ell 4}$  &  $-2.2 \cdot 10^{-6}$ & 
$1.5 \cdot 10^{-8}$ & 
 $-4.4 \cdot 10^{-5}$
&
$3.0 \cdot 10^{-7}$ 
& 
$-2.0 \cdot 10^{-3}$
&  $1.3 \cdot 10^{-5}$ \\
\hline
\hline
&
\multicolumn{2}{c||}{$[C^*_{d u \nu e}]_{11}$} & 
\multicolumn{2}{c||}{$[C^*_{d u \nu e}]_{22}$} &
\multicolumn{2}{c|}{$[C^*_{d u \nu e}]_{33 }$ } 
 \\
\hline
$ [C_{\nu F}]_{\ell 4}$ &
\multicolumn{2}{c||}{$-2.7 \cdot 10^{-13}\, [Y_e]_{\ell \ell}$}  & 
\multicolumn{2}{c||}{$-2.9 \cdot 10^{-9}\, [Y_e]_{\ell \ell}$} & \multicolumn{2}{c|}{$-2.8 \cdot 10^{-5}\, [Y_e]_{\ell \ell}$}   \\
\hline
\hline
&
\multicolumn{2}{c||}{$\left[C_{L\nu Le}\right]_{e4ee}$} & 
\multicolumn{2}{c||}{$\left[C_{L\nu Le}\right]_{\mu4\mu\mu}$} &
\multicolumn{2}{c|}{$\left[C_{L\nu Le}\right]_{\tau4\tau\tau}$ } 
 \\
\hline
$ [C_{\nu F}]_{\ell 4}$ &
\multicolumn{2}{c||}{$- 8.5 \cdot 10^{-8}$}  & 
\multicolumn{2}{c||}{$- 1.3 \cdot 10^{-5}$} & \multicolumn{2}{c|}{$- 1.8 \cdot 10^{-4} $}   \\
\hline
\end{tabular}
\caption{
Neutrino dipole operator, in units of $v^2/\Lambda^2$, at the scale $\mu = 2\,\mathrm{GeV} \,(\mu = m_r)$ as a function of semileptonic (leptonic) $\nu$SMEFT operators at the new physics scale $\Lambda = 10^3\,\mathrm{TeV}$. 
For the semilpetonic operators the subscript denote quark generation indices.
All operators have leptonic indices, $rs$, which we omit.
For operators involving the quark third generation, we integrate out the $b$ quark at $\mu = m_b$.
}
\label{tab:rge_sol}
\end{table}

To show the size of the RG mixing, in Tab.~\ref{tab:rge_sol} we provide the 
coefficient of the dipole operator $C_{\nu F}$
at the scale $\mu_{\rm low}$, choosing as initial scale $\mu_{\rm high} = 10^3\,\mathrm{TeV}$.  
The coefficient is given in the flavor basis.
For the leptonic operator $C_{L\nu L e}$
the low-energy scale $\mu_{\rm low}$ is determined by the mass of the charged lepton. In the case of semileptonic operators, $\mu_{\rm low} = 2\,\mathrm{TeV}$ for operators with only light quarks, while we stop the evolution at $\mu_{\rm low} = m_b$ for scalar and tensor operators with $b$ quarks. 
The pattern emerging in Tab.~\ref{tab:rge_sol} reflects the RGEs in Eqs.~\eqref{eq:nuW}-\eqref{eq:LnuuQ}. The operators with the largest mixing are $C_{L\nu Le}$ and $C^{(3)}_{L\nu Qd}$, which mix directly onto dipoles, proportionally to the Yukawa couplings of charged leptons and $d$-type quarks. The scalar operator $C^{(1)}_{L\nu Qd}$
first mixes into the tensor operator, which then feeds into the dipole. The scalar-tensor mixing is only active between $\Lambda$ and the EW scale, and is driven by electroweak loops. The large logarithm is not sufficient to fully compensate for the loop factor and EW gauge couplings, and thus  
scalar contributions are suppressed by $3 \cdot 10^{-3}$, not too different from a typical EW loop. The right-handed charged-current operator $C_{du\nu e}$ also mixes onto the tensor operator. In this case the mixing is driven by the Yukawa couplings
of charged leptons and $u$-type quarks.
It is thus most important for operators in the third 
generation.  Eq. \eqref{eq:LnuuQ} shows that the semileptonic scalar operator $C_{Q u \nu L}$ could generate a dipole at three loops, by first running into $C_{du \nu e}$,
which runs into the tensor operator $C^{(3)}_{L\nu Qd}$,
which finally runs into $C_{\nu F}$. While this is still formally a leading-log path, the mixing is proportional to three powers of quark Yukawas ($Y^2_d Y_u$)
and two powers of charged lepton Yukawas ($Y^2_e$). 
For light quarks and leptons, $\mathcal O_{Q u L \nu}$ can cancel the dimension-4 contribution to $X$ rays only if $\Lambda < v$, which spoils the $\nu$SMEFT approach.

We point out
a further contribution
to $\nu_R \rightarrow \nu_L \gamma$ transitions, which are induced by nonperturbative effects.
The $\nu$LEFT tensor operator $C_{\nu d}^{T, RR}$, and, by extensions, the $\nu$SMEFT tensor operator
$C^{(3)}_{L\nu Qd}$ and the other operators running into it,
has a non-zero matrix element $\langle \gamma(q) | \bar d \sigma^{\mu \nu} d | 0 \rangle \propto q^\mu \varepsilon^{*\nu} - q^\nu \varepsilon^{*\mu}$, where $\varepsilon$ is the photon polarization.
This has the same form as the matrix element of a dipole operator, and is induced by the correlator of the tensor current with the electromagnetic quark current~\cite{Cata:2007ns,Dekens:2018pbu}.  
The zero momentum vector-tensor correlator as been calculated in Ref.~\cite{Cata:2007ns,Dekens:2018pbu,Cirigliano:2021img}, and leads to a photon dipole of the size as given in Eq.~\eqref{app:eq:np} in App.~\ref{app:np}. 
We find that only for the down quark this non-perturbative contribution can actually dominate over the above discussed perturbative contribution by a factor of $\mathcal{O}(1-10)$.
This contribution would not change the qualitative main features in our analysis, and, being the nonperturbative matrix element affected by sizable theoretical uncertainty, we decided to neglect it in the numerical evaluation of Sec.~\ref{subsubsec:xray_cancellation}.

\subsection{Flavor and mass basis relation}
We finally remark that the RGEs are given in the flavor basis, while the decay rate
discussed in  Sec.~\ref{subsubsec:xray_cancellation}  
is more conveniently expressed in terms of the physical dipole operator,
\begin{equation}
    \mathcal L_{\rm dipole} = \frac{1}{v} 
    C_{\substack{\nu F \\ ij}} \bar \nu_i \sigma^{\mu \nu} P_R \nu_j F_{\mu \nu} + \mathrm{h.c.}\,,
\end{equation}
with $\nu_{i, j}$ 
neutrino mass eigenstates.
The relation is
\begin{align}
\label{eq:Cij_to_Crs}
C_{\substack{\nu F  \\ i j}} = \sum_{r, s}  C_{\substack{\nu F \\ r s}} U^*_{r i} U^{*}_{s j}
    + \sum_{s_1, s_2}\frac{v}{\Lambda}c_w 
    C^R_{\substack{\nu B \\ s_1 s_2}} U^*_{s_2 j} U^*_{s_1 i}
    \,,
\end{align}
where the sum is extended over $r \in \{e, \mu, \tau\}$, while the sterile indices $s$, $s_{1}$ and $s_2$ run over the sterile flavor indices and $\left[C_{\substack{\nu F}}\right]_{rs}$ contains contributions from dimension $4$ and $6$.
For the dimension-$4$ contribution, see Eq.~\eqref{eq:sm}, we express the Dirac mass as
\begin{align}
    \left[m_D\right]_{rs} = \sum_k  m_k U_{r k}  U_{s k}\,,
\end{align}
so that Eq.~\eqref{eq:sm} becomes 
\begin{align}
\label{eq:CnuF_d4}
    C_{\substack {\nu F\\ ij}}= \frac{e }{16 v \pi^2}\sum_{r=e,\mu,\tau}U^*_{ri}U_{rj}F(x_r)m_j\,,
\end{align}
where we used that the mixing matrix in the sterile sector satisfies $\sum_s U_{sk} U^*_{sj} = \delta_{k j}$.
From Eq.~\eqref{eq:CnuF_d4} it is then clear that the dimension-$4$ contribution is suppressed by a factor of $U_{r 4} =\mathcal O(\theta)$.
The contribution of the sterile neutrino Majorana magnetic moment similarly includes a factor of the active-sterile mixing and is also $\mathcal O(\theta)$.
It is however not suppressed by the light sterile neutrino mass or by loop factors, leading to $\Lambda\sim {\cal O}(10^7)\,\mathrm{TeV}$, if the Majorana magnetic moment is to cancel the contribution in Eq.~\eqref{eq:CnuF_d4}. 
It should be noted that the flavor index in the operator should be different, $s_1\neq s_2$.
Our study assumes $\nu$SMEFT operators in Tab.~\ref{tab:nuSMEFT_operators_dim6} only involve the lightest sterile neutrino, so we will ignore the operator $C_{\nu B}^R$ in the rest of the paper. 
A DM scenario that employs the sterile neutrino Majorana magnetic moment is discussed in~\cite{Cho:2021yxk}. 
Dimension-$6$ operators that mix into the Dirac dipole $C_{\nu F}$ via Eqs.~\eqref{eq:nuW}-\eqref{eq:LnuuQ} only involve $\mathcal O(1)$ mixing factors, leading to severe constraints on their scales, as we will see later. Finally, operators that mix via the sterile neutrino Yukawa, as in Eqs.~\eqref{eq:LnuLeY} and \eqref{eq:LnuQd_tensorY}, have the same loop, sterile neutrino mass and active-sterile mixing factors as the dimension-$4$ contribution, and are further suppressed by $m_f^2/\Lambda^2$, where $m_f$ denotes the mass of charged leptons or $d$-type quarks. Cancelling the dimension-$4$ contribution would then require scales well below the EW, beyond the validity of $\nu$SMEFT.


\section{Sterile neutrino as dark matter}
\label{sec:sterile_nu_dm}
In this section we discuss the production mechanism of sterile neutrino as DM.
We review the standard dimension-$4$ mechanism and then show that $\nu$SMEFT interactions can lead to a successful sterile neutrino freeze-in.
The results are confronted to current astrophysical and cosmological constraints and we show how these affect the parameter space to match the observed DM abundance.

\subsection{Sterile neutrino freeze-in via mixing}
\label{subsec:DW}

The dimension-$4$ mixing of active and sterile neutrinos, see Eq.~\eqref{eq:CI_mixing}, inevitably leads to a production of the latter ones in the early Universe.
At temperatures above $1\,\mathrm{MeV}$ active neutrinos are in a state of thermal and (almost perfect) chemical equilibrium.
Their weak interactions with the SM plasma, denoted by the rate $\Gamma_a$, induce a sterile neutrino thermalization rate through oscillations~\cite{Barbieri:1989ti}
\begin{align}
\label{eq:Gamma_s_DW}
    \Gamma^{\mathrm{DW}}_s \simeq \frac{1}{2} \sum_a \langle P\left(\nu_a \to \nu_s \right) \rangle \Gamma_a \simeq \frac{90 \zeta(3)}{7 \pi^4} G_F^2 T^4p \sum_a y_a\,\langle P\left(\nu_a \to \nu_s \right) \rangle \,.
\end{align}
The active neutrino momentum is well approximated by the average of the Fermi-Dirac distribution $p \simeq \langle p \rangle = 3.15 T$, the time-averaged active to sterile neutrino oscillation probability is $\langle P\left(\nu_a \to \nu_s \right) \rangle$ and the flavor diagonal Yukawa couplings are $y_e = 3.6$ and $y_{\mu (\tau)} = 2.5$ for $T \lesssim 20 (180)\,\mathrm{MeV}$ and $y_{\mu (\tau)} = y_e$ for higher temperatures~\cite{Dolgov:1999wv}.
Of course this transition probability in general is a non-trivial function of temperature and momentum of the active neutrino, as well as the mass of the sterile one.
Following Ref.~\cite{Hernandez:2013lza}, simple analytical estimates at leading order in the (small) active-sterile neutrino mixing can be found
\begin{align}
    \langle P\left(\nu_a \to \nu_s \right) \rangle = 2 \left( \frac{\mnus^2}{2 pV_a - \mnus^2} \right)^2 \theta^2 + \mathcal{O}(\theta^4)\,,
\end{align}
with the active neutrino thermal potential 
\begin{align}
    V_a = - f_a  \frac{7 \sqrt{2} \pi^2}{45} G_F \frac{2+c_w^2}{m_W^2} T^4 p\,, 
\end{align}
where $f_a = 1$ at high temperatures for all flavors $a=\{e,\mu,\tau\}$, but $f_{\mu(\tau)} = (1+2 c_w^{-2})^{-1}$ for $T \lesssim 20 (180)\,\mathrm{MeV}$.
Taking the temperature derivative of Eq.~\eqref{eq:Gamma_s_DW}, we see that the production of the sterile neutrinos peaks for
\begin{align}
\label{eq:DW_max_prod_temp}
    T^{\mathrm{DW}}_{\mathrm{max}} \simeq 122 \left( \frac{m_{\nu_s}}{\mathrm{keV}} \right)^{1/3}\,\mathrm{MeV}\,.
\end{align}
For $T \lesssim T^{\mathrm{DW}}_{\mathrm{max}}$ we have $ \Gamma^{\mathrm{DW}}_s \propto T^5$, while in the other limit of $T \gtrsim T^{\mathrm{DW}}_{\mathrm{max}}$ it is $ \Gamma^{\mathrm{DW}}_s \propto T^{-7}$, such that the sterile neutrino production indeed shows a pronounced peak at $T^{\mathrm{DW}}_{\mathrm{max}}$.

\begin{figure}[!t]
\centering
\begin{tabular}{cc}
\hspace{-0.9cm} \includegraphics[width=0.53\textwidth]{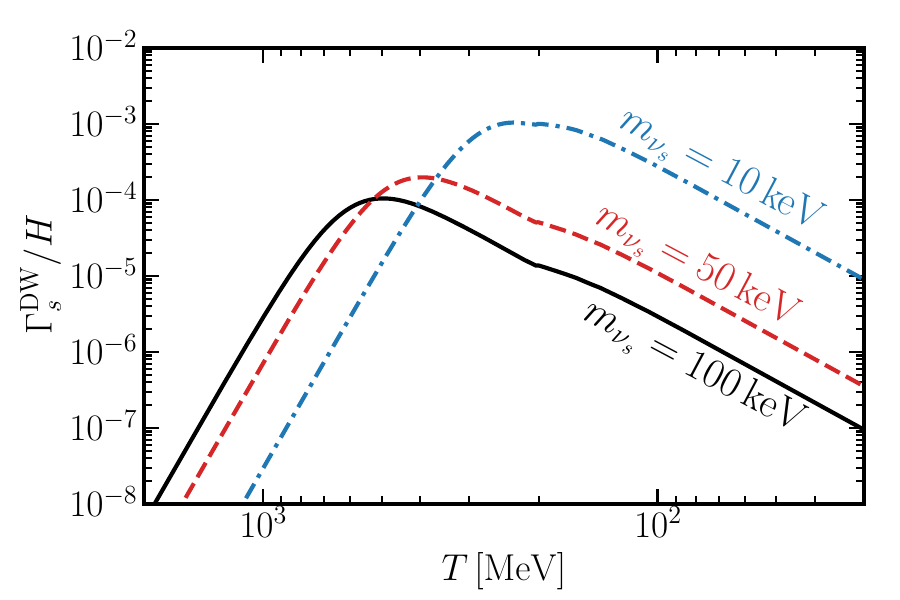} &
\hspace{-0.5cm}  \includegraphics[width=0.53\textwidth]{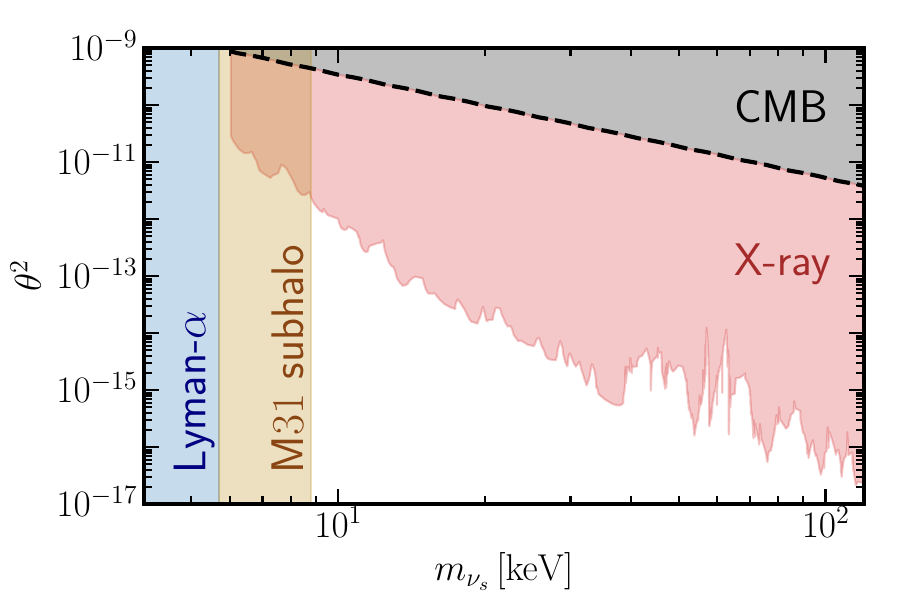}    
\end{tabular}
\vspace{-0.4cm}
\caption{
\textit{Left:}
Production rate normalized by the Hubble expansion for the standard Dodelson-Widrow sterile neutrino production via mixing, see Eq.~\eqref{eq:Gamma_s_DW}.
The mixing for each $\mnus$ is fixed by Eq.~\eqref{eq:DW_contour} to match the observed DM abundance.
\textit{Right:}
Constraints on sterile neutrino from X-ray (red)~\cite{Ng:2019gch, Roach:2022lgo}, Lyman-$\alpha$ (blue)~\cite{Irsic:2023equ} and CMB observations~\cite{Vincent:2014rja}, as well as $\mathrm{M}31$-like N-body simulations (yellow)~\cite{Horiuchi:2013noa}, as a function of the sterile neutrino mass against mixing. 
The black dashed line represents the DW line of Eq.~\eqref{eq:DW_contour}.
}
\label{fig:cons_cosmo}
\end{figure}

The evolution of Eq.~\eqref{eq:Gamma_s_DW} for fixed $\theta$ and $\mnus$, normalized to the Hubble rate of the radiation dominated epoch
\begin{align}
\label{eq:hubble_rad}
    H = \frac{T^2}{m_{\mathrm{pl}}} \sqrt{\frac{4 \pi^3 g_*}{45}}\,,
\end{align}
is shown in the left panel of Fig.~\ref{fig:cons_cosmo}.
Here, $m_{\mathrm{pl}} = 1.26\cdot 10^{19}\,\mathrm{GeV}$ is the Planck mass and $g_* =g_*(T)$ denotes the effective number of relativistic degrees of freedom for the energy density of the thermal plasma~\cite{Husdal:2016haj}.
For the range of temperatures of interest $g_*$ varies between $10.75$ and $106.75$.
For $\Gamma^{\mathrm{DW}}_s  \ll H$ the sterile neutrino does not thermalize with the SM plasma and vice versa.

The relic density of sterile neutrinos arising from the production via mixing is then given by~\cite{Dodelson:1993je}
\begin{align}
      \Omega_{\mathrm{s}} \equiv \frac{\rho_s}{\rho_{\mathrm{crit}}} \simeq 6.5 \cdot 10^7 g_*^{-1/2} \theta^2  \left( \frac{\mnus}{\mathrm{keV}}\right)^{2}\,,
\end{align}
with the measured value $\rho_{\mathrm{crit}} = 3.65 \cdot 10^{-11} \,\mathrm{eV}^4$~\cite{Planck:2018vyg}.
Demanding that $\Omega_{\mathrm{s}} = \Omega_{\mathrm{DM}} = 0.267$~\cite{Planck:2018vyg}, fixes the mixing as a function of the sterile neutrino mass
\begin{align}
\label{eq:DW_contour}
     \theta^2_{\mathrm{DW}} \sim 1.5 \cdot 10^{-8} \Omega_{\mathrm{DM}} \sqrt{g_{*}} \left( \frac{\mathrm{keV}}{\mnus}\right)^{2}\,.
\end{align}
This contour line is known as the Dodelson-Widrow line~\cite{Dodelson:1993je} and is shown as the black dashed line in the right panel of Fig.~\ref{fig:cons_cosmo}.
Mixings larger than those given in Eq.~\eqref{eq:DW_contour} lead to a DM overproduction, whereas smaller mixings to an underproduction.

However, it is known that in the presence of primordial leptonic asymmetries, the sterile neutrino production can be resonantly enhanced and the observed DM density can be explained for mixings much smaller than those of Eq.~\eqref{eq:DW_contour}~\cite{Shi:1998km}.
Such asymmetries can indeed be provided via leptogenesis, driven by the two other sterile neutrinos of the theory, but require a considerable amount of fine tuning in some of the model parameters to reach the necessary level of the leptonic asymmetry~\cite{Asaka:2005pn}.
The available parameter space to explain the DM abundance in this scenario is very limited due to strong $X$-ray and Big Bang Nucleosynthesis constraints~\cite{Ng:2019gch, Roach:2022lgo}. 
Furthermore, it was shown that neutrino non-standard interactions can significantly enhance or suppress the sterile neutrino production~\cite{DeGouvea:2019wpf, Kelly:2020pcy, Astros:2023xhe, Bringmann:2022aim, An:2023mkf}, such that smaller as well as larger mixings than those of Eq.~\eqref{eq:DW_contour} can lead to the DM relic abundance.
We will not consider such model specific extensions, but rather will be interested in the production of sterile neutrino DM via $\nu$SMEFT interactions.
In Sec.~\ref{subsec:DM_EFT} we will explicitly calculate the expected DM relic abundance for an exemplary $\nu$SMEFT operator, but let us first summarize the existing constraints on sterile neutrinos being DM in the following Sec.~\ref{subsec:cosmo_cons}.

\subsection{Constraints on sterile neutrino as dark matter}
\label{subsec:cosmo_cons}

For sterile neutrino DM several constraints exist.
First, notice that sterile neutrinos in general decay into the three light, active neutrinos, and if kinematically allowed into other leptons, quarks and bosons.
However, since DM has been observed today we need to require that the sterile neutrino lifetime obeys $\tau_{\nu_s} \geq t_U = 6.6 \cdot 10^{32}/\mathrm{eV}$, with $t_U$ the age of our Universe~\cite{Planck:2018vyg}.
This leads for the sterile neutrino at the $\mathcal{O}(\mathrm{keV})$ scale to a bound on the mixing
\begin{align}
    \theta^2 \leq 6.7 \cdot 10^{-4} \left( \frac{10\,\mathrm{keV}}{m_{\nu_s}} \right)\,.
\end{align}
Much stronger bounds arise from a variety of cosmological and astrophysical observations and we list the most important ones in the following.

\textbf{CMB.}\,
Sterile neutrinos contribute to the DM density, which is precisely measured by the Planck satellite~\cite{Planck:2018vyg}.
Thus it has to be demanded that the sterile neutrino does not surpasses today's measured DM density.
This sets a bound on the mixing to the active sector for sterile neutrino masses in the range $10\,\mathrm{eV} \lesssim \mnus \lesssim 1\,\mathrm{MeV}$, which corresponds to a bound on the mixing as given in Eq.~\eqref{eq:DW_contour}, see Ref.~\cite{Vincent:2014rja} for a detailed analysis.

\textbf{Lyman-$\alpha$.}\,
Using the tight correlation between the local Lyman-$\alpha$ optical depth and the local DM density, observations of absorption lines from distant quasars can be used to extract information of the DM particle properties.
The most stringent bound arises from free-streaming requirements derived from the Lyman-$\alpha$ matter power spectrum 
at red-shifts of $4.2 \lesssim z \lesssim 5$, which can be translated to an absolute lower bound of $\mnus \gtrsim 5.7\,\mathrm{keV}$ at $95\%$ C.L.~\cite{Irsic:2023equ}.

\textbf{M31 subhalos.}\,
N-body simulations of an Andromeda ($M31$) like galaxy incorporating the effect of a generic warm DM particle constrains $\mnus \gtrsim 8.8 \,\mathrm{keV}$~\cite{Horiuchi:2013noa}.
In short, this is because sterile neutrinos with smaller mass do not have sufficient gravitational binding to explain the number of observed sub-halos.

\textbf{X-ray.}\,
Sterile neutrinos decay via $\nu_s \to \nu_i \gamma$, see left panel of Fig.~\ref{fig:Xray_standard} for the corresponding Feynman diagram.
This decay give rise to a photon emitted with energy $E_\gamma = \mnus/2$.
Such monochromatic photon lines can be targeted by astrophysical experiments.
The non-observation of any excess in the $X$-ray emission in the latest NuSTAR $M31$ observations limits decays of the type $\nu_s \to \nu_i\,\gamma$ to occur at rates less than~\cite{Ng:2019gch, Roach:2022lgo}
\begin{align}
\label{eq:x_ray_decay_rate_bound}
    \Gamma \lesssim 10^{-28}/\mathrm{s}\,,
\end{align}
for masses $6 \lesssim \mnus/\mathrm{keV} \lesssim 200$.
This can be translated into a bound of the active to sterile neutrino mixing via~\cite{Pal:1981rm} 
\begin{align}
    \Gamma \simeq 5\cdot 10^{-32}\,\frac{1}{\mathrm{s}}\, \left( \frac{\theta^2}{10^{-10}}\right) \left( \frac{\mnus}{\mathrm{keV}} \right)^5 \lesssim 10^{-28}/\mathrm{s} \,.
\end{align}

The constraints can be summarized in a plot in the plane $(m_{\nu_s}, \theta^2)$ and are shown in the right panel of Fig.~\ref{fig:cons_cosmo}.
In particular, this excludes the minimal DW production mechanism, see Sec.~\eqref{subsec:DW}, of sterile neutrino as DM.

\subsection{Sterile neutrino freeze-in via EFT interactions}
\label{subsec:DM_EFT}

Additional to the irreducible dimension-$4$ sterile neutrino production via mixing as discussed in Sec.~\ref{subsec:DW}, further production channels via $\nu$SMEFT operators, see Sec.~\ref{sec:eft}, are possible.
The operators of interest in this work, highlighted in Tab.~\ref{tab:nuSMEFT_operators_dim6}, lead to generic $4$-fermion interactions, which can generate the sterile neutrino at tree-level.
The production is thus independent from the active-sterile mixing.
The evolution of the total sterile neutrino density, $n_s \equiv n_{\nu_s} + n_{\bar{\nu}_s}$, via processes of the type $1+2 \leftrightarrow 3+\barparen{\nu}_s$ is governed by the associated Boltzmann equation
\begin{align}
\label{eq:boltzmann_eq_def}
\begin{split}
    \frac{d{n}_s}{dt} + 3Hn_s =  \sum_i &\int \mathrm{d}\Pi_1 \mathrm{d}\Pi_2 \mathrm{d}\Pi_3 \mathrm{d}\Pi_s (2\pi)^4 \delta(p_1 +p_2 - p_3-p_s)  |\mathcal{M}_i|^2    \\
    &\cdot \left[ f_1 f_2 (1-f_3)(1-f_s) - f_3 f_s (1-f_1)(1-f_2) \right]\,,
\end{split}
\end{align}
with $f$ the phase space density, $\mathrm{d}\Pi =  d^3p/(2E(2\pi)^3)$ and $ |\mathcal{M}_i|^2 $ the matrix element summed, but not averaged, over initial and final spins which represent a $\barparen{\nu}_s$ production/annihilation channel.
This equation can be substantially simplified by assuming $f_s \ll f_i = f_i^{\mathrm{eq}} \simeq \exp(-E_i/T)$, that is the sterile neutrino number density is negligible compared to the Maxwell-Boltzmann equilibrium number densities of the SM fermions.
In practice this means that only $1+2 \to 3+\barparen{\nu}_s$ processes are active and thus Eq.~\eqref{eq:boltzmann_eq_def} can be written in the compact form
\begin{align}
\label{eq:boltzmann_eq_nus}
     \frac{d{n}_s}{dt} + 3Hn_s = n_s^{\mathrm{eq}} n_3^{\mathrm{eq}} \langle \sigma v \rangle\,.
\end{align}
Here, we introduce the thermal equilibrium number density
\begin{align}
    n^{\mathrm{eq}} = g \int \frac{\mathrm{d}^3p}{(2\pi)^3} f^{\mathrm{eq}} = g \frac{m^2 T}{2 \pi^2} K_2\left(\frac{m}{T} \right)\,,
\end{align}
where $K_i(x)$ denotes the modified Bessel function of second kind, $g$ the number of internal degrees of freedom and we define the thermally averaged cross section~\cite{Edsjo:1997bg, Hall:2009bx}
\begin{align}
\label{eq:sigma_v_def}
     \langle \sigma v \rangle \equiv \frac{1}{n_s^{\mathrm{eq}} n_3^{\mathrm{eq}}} \frac{T}{512\pi^6} \sum_i \int_{s_{\mathrm{min}}}^\infty \mathrm{d}s\mathrm{d}\Omega\, p_{12} p_{3s} |\mathcal{M}_i|^2 \frac{K_1\left( \frac{\sqrt{s}}{T} \right)}{\sqrt{s}}\,,
\end{align}
with $s$ the Mandelstam variable and 
\begin{align}
    p_{ij} = \frac{\sqrt{s - (m_i + m_j)^2} \sqrt{s - (m_i - m_j)^2}}{2 \sqrt{s}}\,.
\end{align}
This allows to estimate the thermalization efficiency of the sterile neutrinos via 
\begin{align}
\label{eq:gamma_s_def}
    \Gamma_s \equiv n_3^{\mathrm{eq}} \langle \sigma v\rangle \sim T^5/\Lambda^4\,,
\end{align}
where the last step is based on dimensional analysis and holds in the relativistic limit.
As long as $\Gamma_s \ll 1/t \sim H$, with the Hubble rate of Eq.~\eqref{eq:hubble_rad} being the timescale of interest in this problem, the sterile neutrinos do not thermalize and the freeze-in approach ($f_s \ll f_i^{\mathrm{eq}}$) is justified. 

On the other hand, $\Gamma_s \gtrsim H$ indicates that the sterile neutrino reaches thermal equilibrium with the SM plasma and Eq.~\eqref{eq:boltzmann_eq_nus} does not capture the dynamics, but rather Eq.~\eqref{eq:boltzmann_eq_def} has to be solved at the phase space level.
However, since $\Gamma_s/H \propto T^3$, the sterile neutrino can be kept in thermal equilibrium only for a rather short time before it subsequently freezes-out, i.e. $\Gamma_s/H \ll 1$.
In particular, for $\Lambda \gtrsim T_{\mathrm{EW}}$ (required by $\nu$SMEFT consistency) and $\mnus$ $\mathcal{O}(\mathrm{keV})$ the freeze-out would happen while the sterile neutrino is still relativistic. 
Thus, its relic density would be related to the photon number density $n_\gamma$ by $n_s/n_\gamma = \mathcal{O}(0.1)$, where the $\mathcal{O}(0.1)$ factor accounts for the change in the number of relativistic degrees of freedom in the entropy density.
The relic density of the sterile neutrinos would then approximately be $\rho_s \sim \mnus n_\gamma \sim 45 \rho_{\mathrm{crit}} (\mnus/\mathrm{keV})$, which is excluded by CMB observations~\cite{Planck:2018vyg}. 
This sets a bound on $\max(\Gamma_s)$ and thus on $\Lambda/T_r$, with $T_r$ the re-heating temperature of our Universe, as will become clear in the following when solving Eq.~\eqref{eq:boltzmann_eq_nus} explicitly.

\begin{figure}[!t]
\centering
\begin{tabular}{c}
\hspace{-0.3cm} \includegraphics[width=0.96\textwidth]{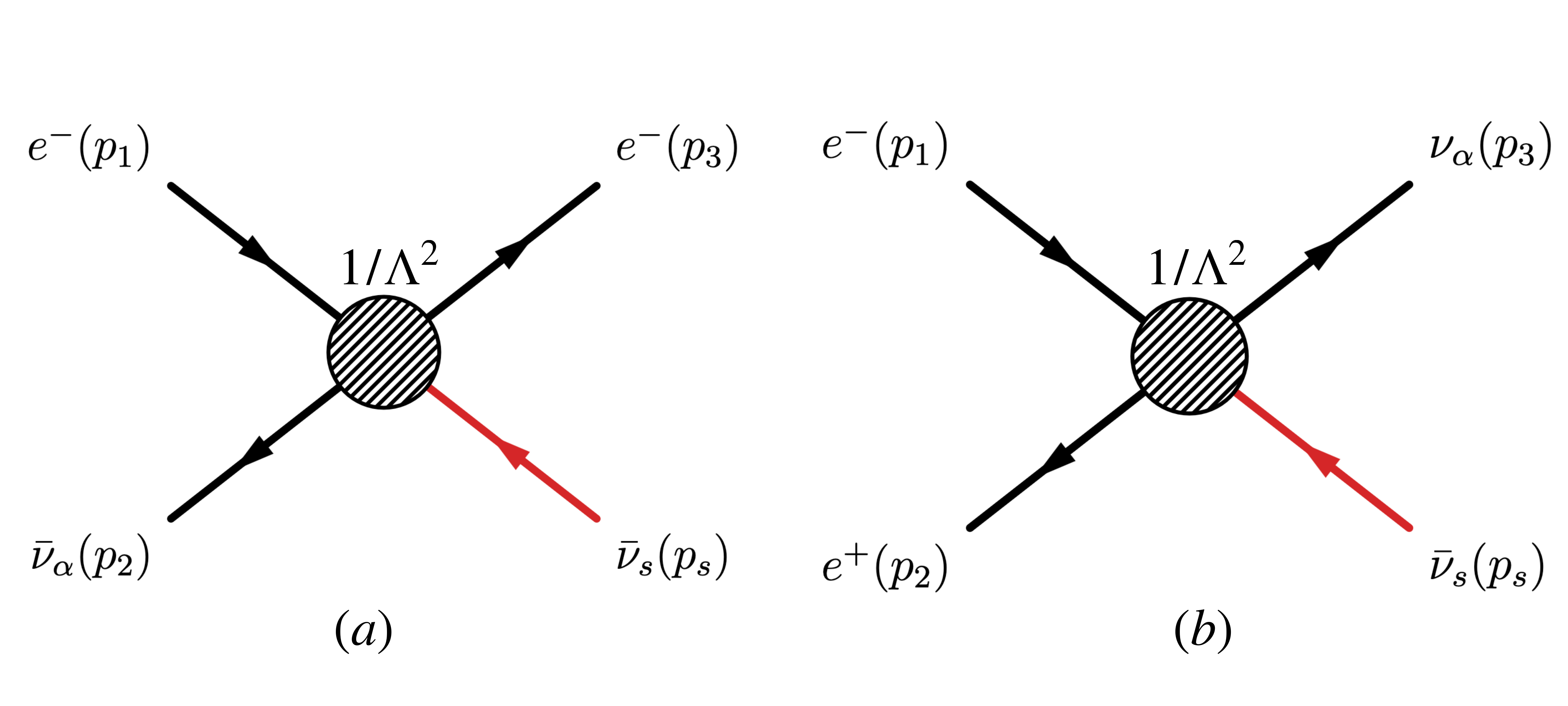} 
\end{tabular}
\vspace{-0.4cm}
\caption{
Feynman diagrams for the $\nu$SMEFT operator $\bar{L}^i \nu_s \epsilon_{ij} \bar{L}^j e$ producing sterile antineutrinos.}
\label{fig:feynman_lnule}
\end{figure}

\begin{figure}[!t]
\centering
\begin{tabular}{cc}
\hspace{-0.95cm} \includegraphics[width=0.525\textwidth]{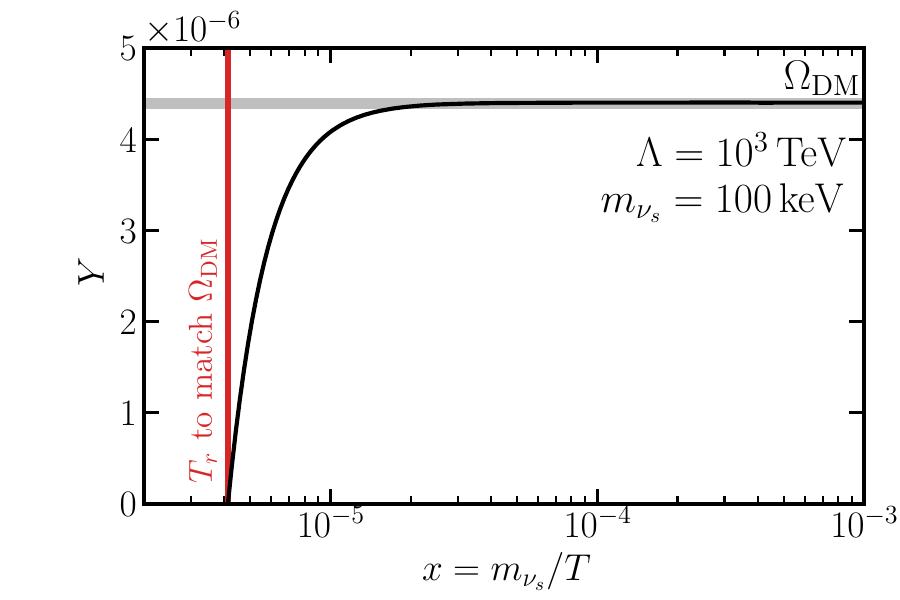} &
\hspace{-0.55cm}  \includegraphics[width=0.525\textwidth]{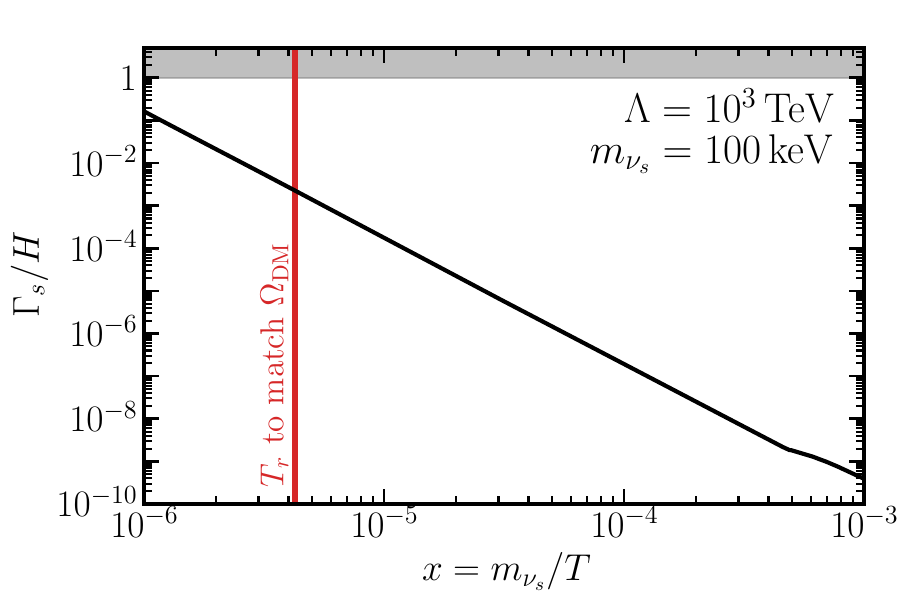}    
\end{tabular}
\vspace{-0.4cm}
\caption{
Sterile neutrino production with the $\bar{L}^i \nu_s \epsilon_{ij} \bar{L}^j e$ operator.
\textit{Left}:
Example of the sterile neutrino freeze-in with $\Lambda = 10^3\,\mathrm{TeV}$ and $\mnus = 100\,\mathrm{keV}$ fixed.
Vertical line indicates the required re-heating temperature $T_r$ to match the observed DM relic abundance.
Larger values for $T_r$ lead to a DM overproduction.
\textit{Right}: The thermally averaged cross section associated to the parameter point of the left panel.
}
\label{fig:freeze_in_example}
\end{figure}

Let us now proceed by solving the Boltzmann equation~\eqref{eq:boltzmann_eq_nus} for $\Gamma_s \ll H$ in the context of the $\nu$SMEFT operators highlighted in Tab.~\ref{tab:nuSMEFT_operators_dim6}.
In order to not spoil the clarity of the presented discussion, we will focus only on the first family of the operator $\bar{L}^i \nu_s \epsilon_{ij} \bar{L}^j e$.
The Feynman diagrams leading to a sterile neutrino production are shown in Fig.~\ref{fig:feynman_lnule}.
For the temperatures of interest, $T \gg \mathcal{O}(\mathrm{MeV})$, the squared matrix elements for topology (a) and (b) are functionally the same and given by 
\begin{align}
\label{eq:M2_lnule}
     |\mathcal{M}|^2  = \frac{s^2}{4 \Lambda^4} (\cos\vartheta - 3)^2\,,
\end{align}
with $\vartheta$ the angle between particle $1$ and $3$.
The hermitian conjugate of the operator $\bar{L}^i \nu_s \epsilon_{ij} \bar{L}^j e$ produces $\nu_s$ and has the same $|\mathcal{M}|^2$ as given in Eq.~\eqref{eq:M2_lnule}.
Hence, for the Boltzmann equations tracking the evolution of $n_s = n_{\nu_s} + n_{\bar{\nu}_s}$ we take twice the amplitude of Eq.~\eqref{eq:M2_lnule}.
The total thermal averaged cross section, Eq.~\eqref{eq:sigma_v_def}, in the relativistic limit is thus
\begin{align}
\label{eq:sigma_v_rel}
    \langle \sigma v \rangle = \frac{7}{2 \pi} \frac{T^2}{\Lambda^4}\,.
\end{align}
To solve for the sterile neutrino number density, Eq.~\eqref{eq:boltzmann_eq_nus}, it is convenient to change variables from $n_s$ and $t$ respectively to the following two parameters:
\begin{align}
\label{eq:Yx_def}
    Y \equiv \frac{n_s}{s} \equiv n_s \left[ \frac{2 \pi^2}{45} g_s T^3\right]^{-1}\,,\,\,\,\,\,\, x \equiv \frac{m_{\nu_s}}{T}\,,
\end{align}
with $g_s = g_s(T)$ counting the number of relativistic degrees of freedom with respect to the entropy density of our Universe~\cite{Husdal:2016haj}.
For the range of temperatures of interest $g_s$ varies between $10.75$ and $106.75$.
With the definitions of Eq.~\eqref{eq:Yx_def} we have 
\begin{align}
\label{eq:final_yield_rel}
    Y &= \int_{x_r=\mnus/T_r}^{\infty} \mathrm{d}x\,  \frac{945 \,\sqrt{5}}{2 \,\pi^{17/2}} \frac{m_{\mathrm{pl}} \mnus^3}{g_s \sqrt{g_*} \Lambda^4} \frac{1}{x^4} \sim \frac{2.6 \cdot 10^{-7}}{g_s \sqrt{g_{*}}} \left( \frac{T_r}{1\,\mathrm{GeV}}\right)^3 \left( \frac{10^3\,\mathrm{TeV}}{\Lambda}\right)^4\,,
\end{align}
with $g_s,\, g_*$ evaluated at $T_r$, with $T_r$ the re-heating temperature acting as an UV cut-off.\footnote{We assume an instantaneous re-heating of our Universe with no assumptions on the inflationary dynamics. In Ref.~\cite{Becker:2023tvd} it was pointed out that for $T_r \ll 100\,\mathrm{GeV}$ the DM production can (mildly) depend on the exact modelling of a non-instantaneous reheating phase.}
An explicit example of the yield $Y$ as well as the normalized sterile neutrino interaction rate is shown in Fig.~\ref{fig:freeze_in_example}. 
The interaction rate shows the anticipated $\Gamma_s/H \propto T^{3}$ behaviour with which we can also easily understand the time evolution of $Y$.
Most of the sterile neutrino production happens around $T_r$ and the asymptotic limit of the final sterile neutrino density is already reach at $T \simeq T_r/8$.
For lower temperatures we see that $\Gamma_s/H \ll 10^{-6}$ and thus no efficient sterile neutrino production is possible anymore.

With the yield of Eq.~\eqref{eq:final_yield_rel} we obtain for the sterile neutrino relic density
\begin{align}
\label{eq:relic_density_rel}
    \Omega_s  \equiv \frac{\rho_s}{\rho_{\mathrm{crit}}} = \frac{\mnus s_0 Y}{\rho_{\mathrm{crit}}} \simeq  \frac{1.5 \cdot 10^{-2}}{g_s \sqrt{g_*}}  \left( \frac{\mnus}{100\,\mathrm{keV}} \right) \left( \frac{10^3\,\mathrm{TeV}}{\Lambda}\right)^4 \left( \frac{T_r}{1\,\mathrm{GeV}} \right)^3\,,
\end{align}
with the entropy density today $s_0 = 2.24\cdot 10^{-11}\,\mathrm{eV}^3$ (assuming a contribution of $N_{\mathrm{eff}} = 3.044$ from the active neutrinos~\cite{Jackson:2023zkl, Drewes:2024wbw}).
To match $\Omega_{\mathrm{s}} = \Omega_{\mathrm{DM}}= 0.267$~\cite{Planck:2018vyg}, the following re-heating temperature for a given $\Lambda$ and $\mnus$ is required
\begin{align}
\label{eq:t_r_match_omega_dm}
    T_r = 2.6 \left( g_s \sqrt{g_*} \right)^{1/3} \left( \frac{100\,\mathrm{keV}}{\mnus}\right)^{1/3} \left( \frac{\Lambda}{10^3\,\mathrm{TeV}} \right)^{4/3}\,\mathrm{GeV}\,.
\end{align}
Note that Eq.~\eqref{eq:relic_density_rel} is indicating that $\Lambda \gg T_r$ is required to explain the observed sterile neutrino DM abundance within the $\nu$SMEFT framework.

Recall, however, that the above discussion is only valid for $\Gamma_s \ll H$.
Combining the result of Eq.~\eqref{eq:sigma_v_rel} with the definition of Eq.~\eqref{eq:gamma_s_def} and demanding the sterile neutrino interaction rate to be smaller than the Hubble rate of Eq.~\eqref{eq:hubble_rad} leads to
\begin{align}
\label{eq:t_reheat_bound}
    T_r \ll 8.4 \cdot 10^{1} g_*^{1/6} \left( \frac{\Lambda}{10^3\,\mathrm{TeV}} \right)^{4/3}\,\mathrm{GeV}\,.
\end{align}
This represents a limit for the discussed sterile neutrino freeze-in scenario.
For larger values of $T_r/\Lambda$ the sterile neutrino would thermalize, but then subsequently freeze-out.
As discussed after Eq.~\eqref{eq:gamma_s_def} such a scenario is cosmological excluded.
On the other hand, the re-heating temperature is constrained from below to be $T_r \gg 5\,\mathrm{MeV}$ in order to not spoil Big Bang Nucleosynthesis~\cite{Hasegawa:2019jsa, Kawasaki:1999na}.
Thus, an absolute lower bound on the EFT scale can be set
\begin{align}
    \Lambda \gg 675\,\mathrm{GeV} {g_*^{-1/8}}\,,
\end{align}
valid as long as we are in the relativistic limit.
When particle masses become important, the bound gets slightly relaxed due to the effective Boltzmann suppression of the particle number densities. 
However, in such a case no analytical approximation is possible and the system has to be solved numerically, with the full matrix elements as presented in App.~\ref{app:M2}.
For the results presented in the following Sec.~\ref{subsec:xray} we will always solve the Boltzmann equation numerically.

In conclusion, we have shown that $4$-fermion $\nu$SMEFT operators can produce sterile neutrinos with an abundance matching that of the observed DM density.
Because such generic interactions are tree-level, no mixing between active and sterile neutrinos are required to efficiently produce the latter one.
Also note that interference effects between the irreducible dimension-$4$ production of the sterile neutrinos via the DW mechanism, discussed in Sec.~\ref{subsec:DW}, and the $\nu$SMEFT production are negligible for the cases of our interest.
Only if the following two conditions are simultaneously fulfilled, sizable interference effects in the sterile neutrino production can appear:
i) The active-sterile neutrino mixing is close to the value given in Eq.~\eqref{eq:DW_contour} and ii) the re-heating temperature (closely) coincides with the temperature of maximal production of the DW mechanism as given in Eq.~\eqref{eq:DW_max_prod_temp}. 
As we will show Sec.~\ref{subsubsec:xray_cancellation}, condition ii) is never fulfilled if the $X$-ray bound is to be avoided and thus we do not have to consider this special case here.

\subsection{$X$-ray emission}
\label{subsec:xray}

\begin{figure}[!t]
\centering
\begin{tabular}{cc}
\hspace{-0.5cm} \includegraphics[width=0.5\textwidth]{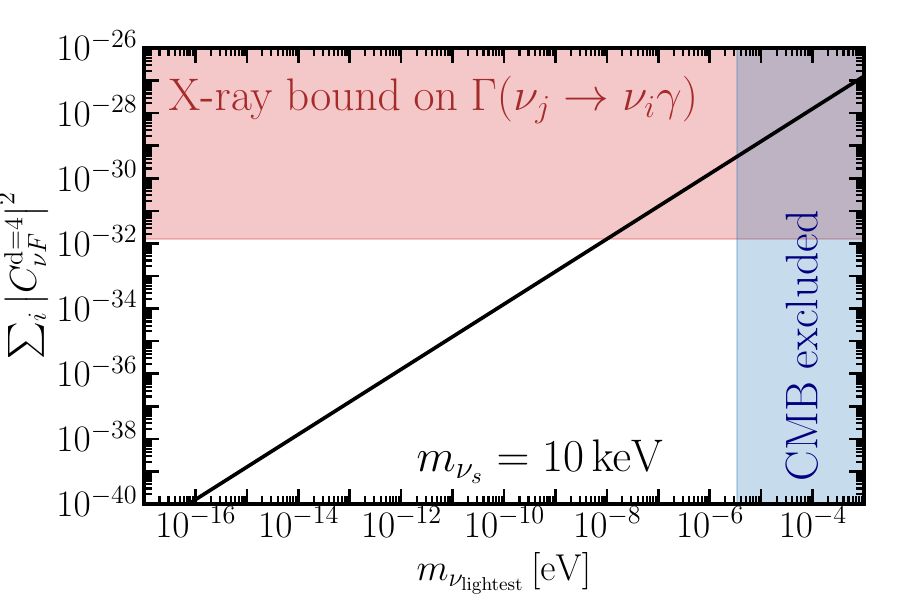} &
\hspace{-0.55cm}  \includegraphics[width=0.5\textwidth]{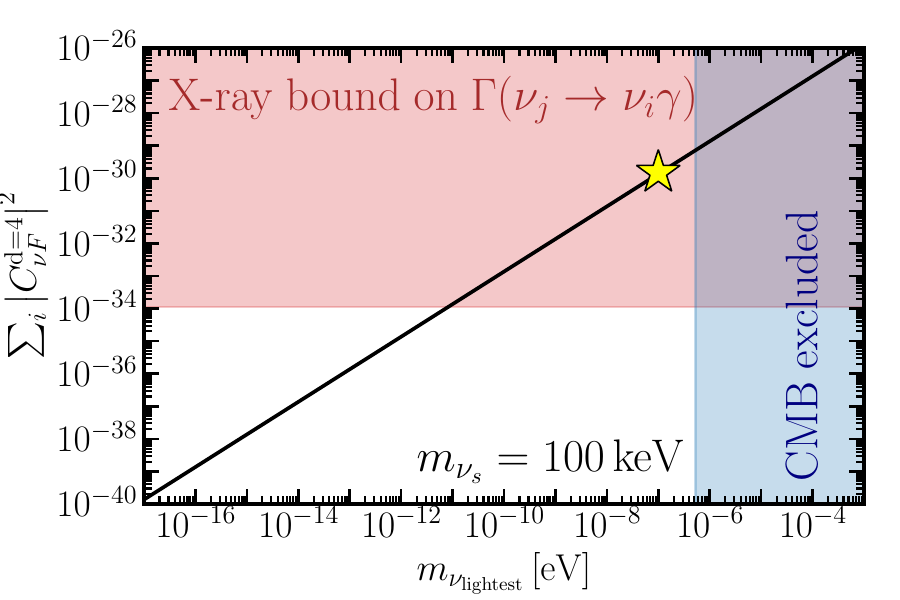}    
\end{tabular}
\vspace{-0.4cm}
\caption{
The black line represents the dipole operator at dimension-$4$ of Eq.~\eqref{eq:sm} for sterile neutrino masses of $\mnus = 10\,(100)\,\mathrm{keV}$ in the left (right) panel.
Colored regions are excluded from $X$-ray observations (red) and from CMB observations (blue), see Sec.~\ref{subsec:cosmo_cons}.
The yellow star indicates the benchmark value considered in Sec.~\ref{subsubsec:xray_cancellation}.
}
\label{fig:C_nuF_SM}
\end{figure}

The strongest bound on the active-sterile neutrino mixing $\theta$ for $\mnus$ $\mathcal{O}(\mathrm{keV})$ arises from non-observations of peaked photon emissions from $M31$~\cite{Ng:2019gch}, see the discussion in Sec.~\ref{subsec:cosmo_cons} and Fig.~\ref{fig:cons_cosmo}.
The decay rate of the process $\nu_j \to \nu_i \gamma$, with $j=4$ and $i=\{1,2,3\}$ mass indices, can be generically expressed via its associated dipole operator
\begin{align}
\label{eq:sterile_nu_radiative_decay_def}
    \Gamma_{\mathrm{tot}} =\sum_{i}\Gamma(\nu_4 \to \nu_i \gamma) =  \frac{\mnus^3}{2 \pi v^2} \sum_{i}  \left\lvert C_{\substack{\nu F \\ i 4}} \right\rvert^2\,.
\end{align}
The dimension-$4$ contribution to the radiative decay, see left panel of Fig.~\ref{fig:Xray_standard}, is given by Eq.~\eqref{eq:sm} and we show the result for fixed $\mnus$ as a function of the lightest active neutrino mass in Fig.~\ref{fig:C_nuF_SM}.
Note that the result are the same for both light neutrino hierarchies. 
In the same figure we also show the bounds adapted from Fig.~\ref{fig:cons_cosmo}, using the relation between the active-sterile neutrino mixing and the lightest neutrino mass from Eq.~\eqref{eq:CI_mixing}.
The strongest bound on the sterile neutrino DM arises from the $X$-ray observations and effectively bounds the dipole operator $C_{\substack{\nu F}}$.
Considering that the sterile neutrino only decays through the dimension-$4$ mixing, we are confined to be on the black line of Fig.~\ref{fig:C_nuF_SM}.

In the general $\nu$SMEFT context, however, the operators discussed in Sec.~\ref{sec:eft} and highlighted in Tab.~\ref{tab:nuSMEFT_operators_dim6} can also contribute to Eq.~\eqref{eq:sterile_nu_radiative_decay_def} and we express the decay rate as
\begin{align}
\label{eq:sterile_nu_radiative_decay_nusmeft}
    \sum_{i}\Gamma(\nu_4 \to \nu_i \gamma) =  \frac{\mnus^3}{2 \pi v^2} \sum_{i}  \left\lvert \sum_X C^X_{\substack{\nu F \\ i 4}} \right\rvert^2\,,
\end{align}
where $X$ indicates the different partial contributions of the photon dipole operator.
Two immediate conclusions from Eq.~\eqref{eq:sterile_nu_radiative_decay_nusmeft} can be derived.
i) Destructive interference effects can suppress the total decay rate $\Gamma$ even though each partial contribution $C^X_{\substack{\nu F}}$ can be large.
ii) Significant contributions to $X$-ray decays can arise even for $\theta = 0$, which in the standard picture escapes detection completely.
Thus, we are not anymore confined to be on the black line of Fig.~\ref{fig:C_nuF_SM}, but rather the whole two dimensional parameter space is now accessible, while still having the sterile neutrino to be the full DM constituent.
In the following subsections we will discuss both cases separately and systematically derive the corresponding new physics scale $\Lambda$ and the re-heating temperature which leads to the observed DM relic abundance.

For the analysis we fix the light neutrino mixing angles to their best-fit values~\cite{Esteban:2020cvm} and consider only the limit of $R = 1$ in Eq.~\eqref{eq:CI_mixing}.
We also fix the Dirac CP phase to the best-fit value~\cite{Esteban:2020cvm}, and leave the two Majorana phases in the $\pmns$ matrix as free parameters allowing them to vary between $0$ and $2\pi$.

\subsubsection{Destructive interference effects and the zero mixing limit}
\label{subsubsec:xray_cancellation}

The operators which can lead to a significant creation of a neutrino photon dipole are highlighted in Tab.~\ref{tab:nuSMEFT_operators_dim6}.
The corresponding RGEs are summarized in Eqs.~\eqref{eq:nuW}-\eqref{eq:LnuQd_scalar} and for the operator matching of $\mathcal{O}_{H \nu e}$ the photon dipole is given in Eq.~\eqref{eq:RH}, while its running is given by Eq.~\eqref{eq:CHnue_rge}.
The operators given in Eqs.~\eqref{eq:LnuLeY}-\eqref{eq:LnuQd_tensorY} and Eq.~\eqref{eq:LnuuQ} also induce a  photon dipole, but can not be used to cancel the $X$-ray emission.
This is because the additional suppression arising from neutrino and charged lepton Yukawas, respectively, lowers the required scales $\Lambda$ for successful $X$-ray cancellation to values below the EW scale.
Such values, however, violate the consistency condition of our $\nu$SMEFT approach and the operators are thus not viable candidates to avoid the $X$-ray bound.

\textbf{Destructive interference.}\,
For a successful cancellation of the $X$-ray emission we demand from Eq.~\eqref{eq:sterile_nu_radiative_decay_nusmeft} that 
\begin{align}
\label{eq:C_nuF_d4_d6}
    \sum_{i} \left\lvert C^{d=4}_{\substack{\nu F \\ i 4}} + C^{d=6}_{\substack{\nu F \\ i 4}}\right \rvert^2= 0 \,.
\end{align}
Since the RGEs are expressed in the flavor basis, it is convenient to evaluate Eq.~\eqref{eq:C_nuF_d4_d6} in the same basis.
Further, we have the freedom to express the Wilson coefficient of a given operator as $\mathcal{C} = |\mathcal{C}| \exp(i \varphi)$ with $\varphi = \mathrm{arg}(\mathcal{C})$.
In fact we can also choose $|\mathcal{C}| = 1$ since we can only constrain $\mathcal{C}/\Lambda^d$ and $|\mathcal{C}| \neq 1$ just corresponds to a shift in $\Lambda$.
Hence, there are only two free parameters $(\Lambda, \varphi)$ for each flavor.
Using Eq.~\eqref{eq:Cij_to_Crs} and Eq.~\eqref{eq:CnuF_d4} we express Eq.~\eqref{eq:C_nuF_d4_d6} as
\begin{align}
\label{eq:C_nuF_d4_d6_flavor}
    \sum_{i} \left\lvert \sum_r U^*_{r i} \left(\frac{e}{16 v \pi^2} i U_{r \ell} \sqrt{\frac{\mnu}{\mnus}} F(x_r) \mnus +  e^{i \varphi_r} \left\lvert C^{d=6}_{\substack{\nu F \\ r s}}(\Lambda_r)\right\rvert  \right)\right \rvert^2= 0 \,,
\end{align}
with $F(x_r)$ defined in Eq.~\eqref{eq:Fxr}, $C^{d=6}_{\substack{\nu F}}(\Lambda_r) \in {\rm I\!R}$, $s$ the flavor index of the sterile neutrino $\nu_4$ and $\ell= 1(3)$ for NH(IH).
Because each individual term of the sum over $i$ is non-negative, we need to ensure that each summand is zero, so Eq.~\eqref{eq:C_nuF_d4_d6_flavor} is equivalent to
\begin{align}
\label{eq:C_nuF_d4_d6_flavor_reduced}
     \sum_r U^*_{r i} \left(\frac{e}{16 v \pi^2} i U_{r \ell} \sqrt{\frac{\mnu}{\mnus}} F(x_r) \mnus +  e^{i \varphi_r} \left \lvert C^{d=6}_{\substack{\nu F \\ r s}}(\Lambda_r) \right \rvert  \right)= 0 \,.
\end{align}
Using the (approximate) unitarity of the PMNS mixing matrix we can multiply both sides by $\sum_i U_{i r'}$ to perform the sum over $i$ and $r$.
For each flavor $r \in \{e,\mu,\tau\}$ the scale $\Lambda_r$ and phase $\varphi_r$ are then given by
\begin{align}
\label{eq:cancellation_sol}
    C^{d=6}_{\substack{\nu F \\ r s}}(\Lambda_r) \simeq \frac{3e}{32 v \pi^2} \sqrt{\mnu\mnus} |U_{r \ell}| \,,\;\;\;\;\; \varphi_r = \arg(U_{r \ell} + \pi/2)\,,
\end{align}
where we used $F(x_r)=-3/2$.
We can see that the phases $\varphi_r$ are aligned with the phases of the PMNS matrix.
On the other hand, to explicitly find $\Lambda_r$ the RGEs of Eqs.~\eqref{eq:nuW}-\eqref{eq:LnuQd_scalar} need to be solved.
For the operator matching of $\mathcal{O}_{H \nu e}$ we can use Eq.~\eqref{eq:RH} to find $\Lambda_r$.\footnote{We remark that the running of $\mathcal{O}_{H \nu e}$ is only a $\mathcal{O}(0.1\%)$ effect on $\Lambda_r$ for $X$-ray cancellation and can be safely neglected.}

The obtained values for $\Lambda_r$ and $\varphi_r$ are summarized for each light neutrino mass ordering, NH and IH, in Tab.~\ref{tab:Xray_scale} for $m_{\nu_{\mathrm{lightest}}} = 10^{-7}\,\mathrm{eV}$ and $\mnus = 100\,\mathrm{keV}$.
\begin{table}[!t]
    \centering
    \begin{tabular}{| c | c || c | c | c || c |}
    \hline
          & $\mathcal{O}_X$ & $\Lambda_e/\mathrm{TeV}$ & $\Lambda_\mu/\mathrm{TeV}$ & $ \Lambda_\tau/\mathrm{TeV}$ & $T_r/\mathrm{GeV}$\\
    \hline
         \multirow{5}{*}{NH} & $\mathcal{O}_{L \nu L e}$ & $2.36 \cdot 10^3$ & $4.88 \cdot 10^4$ & $1.77 \cdot 10^5$ & $55.4$ \\
         & $\mathcal{O}_{L\nu Qd}^{(3)}$ & $1.26 \cdot 10^4$ & $9.36 \cdot 10^4$ & $5.83 \cdot 10^5$ & $119.8$ \\
         & $\mathcal{O}_{L\nu Qd}^{(1)}$ & $9.6 \cdot 10^2$ & $7.97 \cdot 10^3$ & $5.39 \cdot 10^4$ & $13.0$ \\
         & $\mathcal{O}_{d u \nu e}$ & $6.11 \cdot 10^1$ & $1.82 \cdot 10^3$ & $6.83 \cdot 10^3$ & $0.3$ \\
         & $\mathcal{O}_{H \nu e}$ & $5.0 \cdot 10^2$ & $1.1 \cdot 10^4$ & $4.0 \cdot 10^4$ & $5.5$ \\
    \hline
    \hline
         \multirow{5}{*}{IH} & $\mathcal{O}_{L \nu L e}$ & $5.67 \cdot 10^3$ & $3.37 \cdot 10^4$ & $1.43 \cdot 10^5$ & $230.5$ \\
         & $\mathcal{O}_{L\nu Qd}^{(3)}$ & $3.03 \cdot 10^4$ & $6.15 \cdot 10^4$ & $4.89 \cdot 10^5$  & $379.7$ \\
         & $\mathcal{O}_{L\nu Qd}^{(1)}$ & $2.43 \cdot 10^3$ & $5.13 \cdot 10^3$ & $4.48 \cdot 10^4$ & $46.2$ \\
         & $\mathcal{O}_{d u \nu e}$ & $1.57 \cdot 10^2$ & $1.17 \cdot 10^3$ & $5.68 \cdot 10^3$  & $1.2$ \\
         & $\mathcal{O}_{H \nu e}$ & $1.2 \cdot 10^3$ & $7.6 \cdot 10^3$ & $3.3 \cdot 10^4$ & $17.4$ \\
    \hline
    \end{tabular}
    \caption{
    $\nu$SMEFT scales $\Lambda_r$ leading to successful $X$-ray cancellation for $m_{\nu_{\mathrm{lightest}}} = 10^{-7}\,\mathrm{eV}$ and $\mnus = 100\,\mathrm{keV}$, corresponding to the yellow star in Fig.~\ref{fig:C_nuF_SM} if only dimension-$4$ interaction were considered.
    The phases $\varphi_r$ are given in Eq.~\eqref{eq:cancellation_sol}.
    For $\mathcal{O}_{L\nu Qd}$ we consider lepton and quarks of the same generation.  
    For $\mathcal{O}_{du\nu e}$ we consider the lepton coupling to the third generation quarks only, since otherwise the suppression of the $X$-ray emission requires $\Lambda_r \ll v$,
    which is not viable.
    The right column indicates the re-heating temperature $T_r$ for which the model can produce the observed DM relic abundance, which is always $T_r \ll \Lambda_r$.
    }
    \label{tab:Xray_scale}
\end{table}
Generalizing the results of the scales $\Lambda_r$ to different $(m_{\nu_{\mathrm{lightest}}},\, \mnus)$ is achieved via the approximate re-scaling
\begin{align}
\label{eq:Xray_rescaling}
    \Lambda_r \mapsto \Lambda_r \left(\frac{10^{-7}\,\mathrm{eV}}{m_{\nu_{\mathrm{lightest}}}}\right)^{1/4} \left(\frac{100\,\mathrm{keV}}{\mnus}\right)^{1/4} \,,
\end{align}
which holds as long as the $X$-ray contribution at $d=4$ is larger than the current bound on $\Gamma_{\mathrm{tot}}$.

With the scales given in Tab.~\ref{tab:Xray_scale} and phases as in Eq.~\eqref{eq:cancellation_sol}, the benchmark point highlighted with a yellow star in Fig.~\ref{fig:C_nuF_SM} evades the current $X$-ray constraint.
Using Eq.~\eqref{eq:Xray_rescaling} it is then straightforward to find the appropriate scales $\Lambda_r$ for which the $X$-ray constraint can be evaded for any given $(m_{\nu_{\mathrm{lightest}}},\, \mnus)$.
This allows to open the parameter space for sterile neutrino DM for active-sterile neutrinos mixing of up to four orders of magnitude larger than naively expected, see Fig.~\ref{fig:cons_cosmo}.

The DM production proceeds via the $\nu$SMEFT interactions as discussed in Sec.~\ref{sec:sterile_nu_dm}.
Successful DM production fixes $\Lambda/T_r$ and because $\Lambda_r$ is fixed by the $X$-ray cancellation condition, we can fix the re-heating temperature to a unique value.
In the relativistic limit the re-heating temperature is given by Eq.~\eqref{eq:t_r_match_omega_dm} with $\Lambda$ in that equation being the smallest of the $\Lambda_r$ derived from the $X$-ray cancellation condition.
However, in general, we have to take the fermion masses explicitly into account in the DM production. 
That is because for the scales given in Tab.~\ref{tab:Xray_scale} the Boltzmann suppression can be sizable and thus larger $T_r$ are required to match the DM relic abundance than expected from Eq.~\eqref{eq:t_r_match_omega_dm}.
To find the corresponding $T_r$ we solve Eq.~\eqref{eq:boltzmann_eq_nus} numerically and demand that the relic density matches the observed DM abundance within $1\%$.
The results are summarized in the right column of Tab.~\ref{tab:Xray_scale}.
For completeness, we outline the matrix elements for all operators in App.~\ref{app:M2}.

One may note that in Eq.~\eqref{eq:C_nuF_d4_d6} we have been rather radical by demanding an exact cancellation of the $X$-ray emission.
In a conservative fashion we could lift the zero to the current observational bound on the $X$-ray emission.
This would affect the above result the closer $\mathcal{C}_{\nu F}^{\mathrm{d}=4}$ is to the current $X$-ray bound.
In particular, this relaxes the alignment of $\varphi_r$ with the PMNS phases as given in Eq.~\eqref{eq:cancellation_sol} and in principle allow for an arbitrary value of the phase of $\varphi_r$, depending on the concrete numerical value of the bound.
However, in the most interesting region, the large mixing regime, $\varphi_r$ will always be aligned with the PMNS phases.
The scales $\Lambda_r$, however, would merely change and we find that they can be maximally lowered by $1\%$. 

Let us further comment on the generation of a left-handed Majorana mass $m_L$ arising from the $\nu$SMEFT interactions, see Eq.~\eqref{eq:mL_induced}.
Demanding that the $X$-ray constraint should be evaded, effectively fixes $C_{\nu F}$ for a given pair of $(\mnu,\mnus)$.
From Fig.~\ref{fig:C_nuF_SM} we see that $|C_{\nu F}| \ll 10^{-16}$ for successful $X$-ray cancellation.
Thus, with the corresponding solutions of $C_{\nu W}$ and $C_{\nu B}$ of the RGEs of Eqs.~\eqref{eq:nuW}-\eqref{eq:LnuQd_scalar} we find $m_L \ll 10^{-19}\,\mathrm{eV}$.
It is thus too small to be of any importance in the present context, but could be of relevance if no $X$-ray cancellation is demanded.
This scenario and its possible experimental signatures will be studied elsewhere.

\textbf{Zero mixing limit.}\,
If $m_{\nu_{\mathrm{lightest}}} \to 0$ the $X$-ray emission from the dimension-$4$ interaction is negligible.\footnote{We note that $\mnu$ will never be exactly zero due to the $\nu$SMEFT induced $m_L$. This, however, should not be of our concern in the present analysis, since it is always small as shown above.}
In such a scenario, the only photon emission from a radiative neutrino decay can arise from the higher dimensional operators, corresponding to only the second term in Eq.~\eqref{eq:C_nuF_d4_d6}.
There is thus a lower limit on $\Lambda_r$ for each operator which would satisfy current constraints.
The resulting scales, evaluated for $\mnus = 100\,\mathrm{keV}$, are summarized for NH and IH in Tab.~\ref{tab:Xray_scale_no_mixing}.
\begin{table}[!t]
    \centering
    \begin{tabular}{| c | c || c | c | c || c |}
    \hline
          & $\mathcal{O}_X$ & $\Lambda_e/\mathrm{TeV}$ & $\Lambda_\mu/\mathrm{TeV}$ & $ \Lambda_\tau/\mathrm{TeV}$ & $T_r/\mathrm{GeV}$\\
    \hline
         \multirow{5}{*}{NH} & $\mathcal{O}_{L \nu L e}$ & $2.5 \cdot 10^4$ & $5.2 \cdot 10^5$ & $1.9 \cdot 10^6$ & $1.3\cdot 10^3$ \\
         & $\mathcal{O}_{L\nu Qd}^{(3)}$ & $1.3 \cdot 10^5$ & $9.4 \cdot 10^5$ & $6.3 \cdot 10^6$ & $2.7 \cdot 10^3$ \\
         & $\mathcal{O}_{L\nu Qd}^{(1)}$ & $1.2 \cdot 10^4$ & $8.8 \cdot 10^4$ & $6.3 \cdot 10^5$ & $3.9 \cdot 10^2$ \\
         & $\mathcal{O}_{d u \nu e}$ & $7.5 \cdot 10^2$ & $2.0 \cdot 10^4$ & $8.0 \cdot 10^4$ & $9.9$ \\
         & $\mathcal{O}_{H \nu e}$ & $5.0 \cdot 10^3$ & $1.1 \cdot 10^5$ & $4.0 \cdot 10^5$ & $\sim T_{\mathrm{EW}}$ \\
    \hline
    \hline
         \multirow{5}{*}{IH} & $\mathcal{O}_{L \nu L e}$ & $6.0 \cdot 10^4$ & $3.6 \cdot 10^5$ & $1.5 \cdot 10^6$ & $4.3 \cdot 10^3$ \\
         & $\mathcal{O}_{L\nu Qd}^{(3)}$ & $3.2 \cdot 10^5$ & $6.7 \cdot 10^5$ & $5.1 \cdot 10^6$  & $4.2 \cdot 10^3$ \\
         & $\mathcal{O}_{L\nu Qd}^{(1)}$ & $2.9 \cdot 10^4$ & $6.0 \cdot 10^4$ & $5.1 \cdot 10^5$ & $1.3 \cdot 10^3$ \\
         & $\mathcal{O}_{d u \nu e}$ & $1.9 \cdot 10^3$ & $1.4 \cdot 10^4$ & $6.5 \cdot 10^4$  & $36$ \\
         & $\mathcal{O}_{H \nu e}$ & $1.2 \cdot 10^4$ & $7.6 \cdot 10^4$ & $3.3 \cdot 10^5$ & $\sim T_{\mathrm{EW}}$ \\
    \hline
    \end{tabular}
    \caption{
    $\nu$SMEFT scales $\Lambda_r$ which can lead to a sterile neutrino DM production in the zero mixing limit and escape detection from $X$-ray emission lines. 
    Sterile neutrino mass fixed to $\mnus = 100\,\mathrm{keV}$.
    Flavor assignment as in Tab.~\ref{tab:Xray_scale}.
    The re-heating temperature for $\mathcal{O}_{H \nu e}$ will be around $T_{\mathrm{EW}}$ and a more careful treatment of the freeze-in dynamics is necessary, which is beyond the scope of the paper, although we remark that the DM abundance can certainly be explained with this operator.
    }
    \label{tab:Xray_scale_no_mixing}
\end{table}
For different sterile neutrino masses the scales are found via the replacement $\Lambda_r \mapsto \Lambda_r (100\,\mathrm{keV}/\mnus)^{1/4}$.

So far we used current $X$-ray observations to constrain the scale of $\nu$SMEFT operators.
On the other hand, there are a number of $\nu$SMEFT operators which do not lead to a significant photon dipole, see the unmarked operators in Tab.~\ref{tab:nuSMEFT_operators_dim6}.
These operators, however, can still be effective portals for the sterile neutrino DM production via the very similar $4$-fermion processes as described in Sec.~\ref{subsec:DM_EFT}.
Thus, for active-sterile neutrino mixings which fall below the current constraint imposed by $X$-ray observations, and hence far below the DW line, see Eq.~\eqref{eq:DW_contour}, these operators can still produce the observed DM abundance.


\section{Laboratory searches}
\label{sec:lab}
In order not to violate the $X$-ray bounds,
the scale of the $\nu$SMEFT operators that mix onto the photon dipole needs to be very high, thousands of TeVs for the scalar/tensor four-fermion operators $\mathcal O_{L\nu Le}$, $\mathcal O^{(1,3)}_{L\nu Qd}$ and hundreds of TeVs for vector interactions such as $\mathcal O_{du\nu e}$
and $\mathcal O_{H\nu e}$, see Tab.~\ref{tab:Xray_scale}.
While these scales are very high, it is worth discussing whether they can be probed in present and future laboratory experiments. 
Between the most sensitive probes of BSM physics we can consider searches for neutrinoless double beta decay, precision beta decays of mesons, and electron dipole moments. 

\begin{figure}[!t]
\centering
\begin{tabular}{cc}
\hspace{-0.6cm}\includegraphics[width=0.75\textwidth]{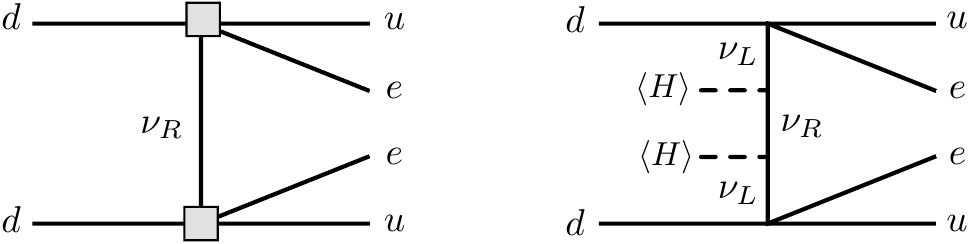}
\end{tabular}
\caption{$0\nu\beta\beta$ diagrams generated by dimension 6 operators (left) and dimension 4 Yukawa interactions (right). $\langle H \rangle$ represents the Higgs vacuum expectation value. A source of lepton-number violation stems from the Majorana mass of the sterile neutrinos in both cases. }
\label{fig:feynman_DBD}
\end{figure}

$\nu$SMEFT operators that give charged currents with first-generation quarks contribute to neutrinoless double beta decay ($0\nu\beta\beta$) in addition to the dimension $4$ interactions. 
The left diagram in Fig.~\ref{fig:feynman_DBD} depicts a $0\nu\beta\beta$ process from dimension 6 operators, in which the (virtual) sterile neutrino line is closed without any suppression of active-sterile mixing parameters. 
The right diagram corresponds to the standard process originating from the Yukawa interactions at dimension $4$.
Following the $\nu$SMEFT analysis of Ref.~\cite{Dekens:2020ttz} we can express the half-life of neutrinoless double beta decay as
\begin{align}
\label{eq:tnu}
    (\tnu)^{-1} = g_A^4 G_{01} \left \lvert \sum_{i=1}^{6} V_{ud}^2 \frac{U_{ei}^2 m_i}{m_e} A_\nu^{\mathrm{d}=4} + A_\nu^{\mathrm{d}=6} \right \rvert^2\,,
\end{align}
in which $g_A = 1.2754 \pm 0.0013$~\cite{ParticleDataGroup:2022pth}, $V_{ud} \simeq 0.97$ the $u$-$d$ CKM element~\cite{ParticleDataGroup:2022pth} and $G_{01} = 15\,(2.4)\cdot 10^{-15}\,\mathrm{yr}^{-1}$ a phase space factor for $^{136}$Xe ($^{76}$Ge)~\cite{Kotila:2012zza}.
The amplitudes $A_\nu$ are calculated using the formulas given in Ref.~\cite{Dekens:2020ttz} and contain long-range as well as short-range nuclear effects.

\begin{figure}[!t]
\centering
\begin{tabular}{cc}
\hspace{-0.6cm} \includegraphics[width=0.51\textwidth]{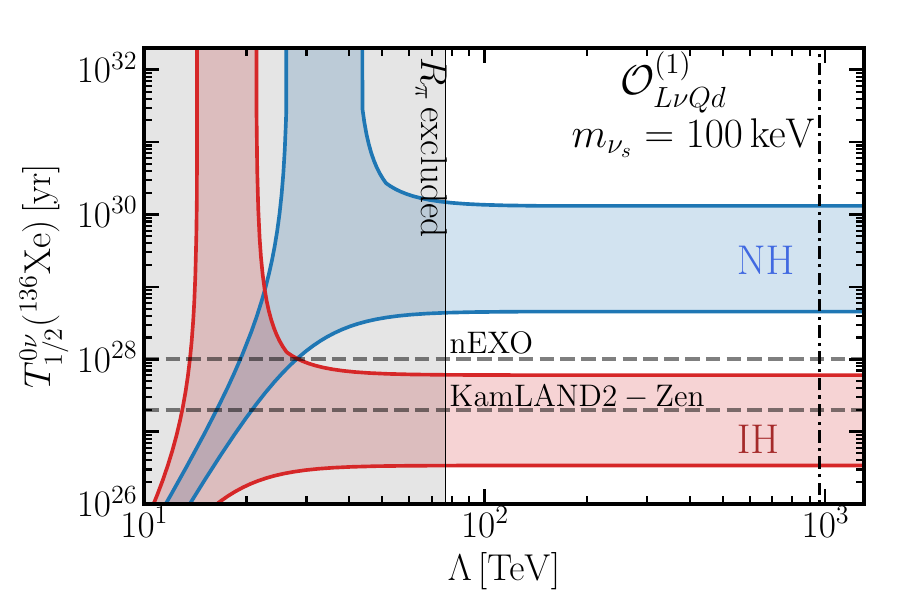} &
\hspace{-0.5cm}  \includegraphics[width=0.51\textwidth]{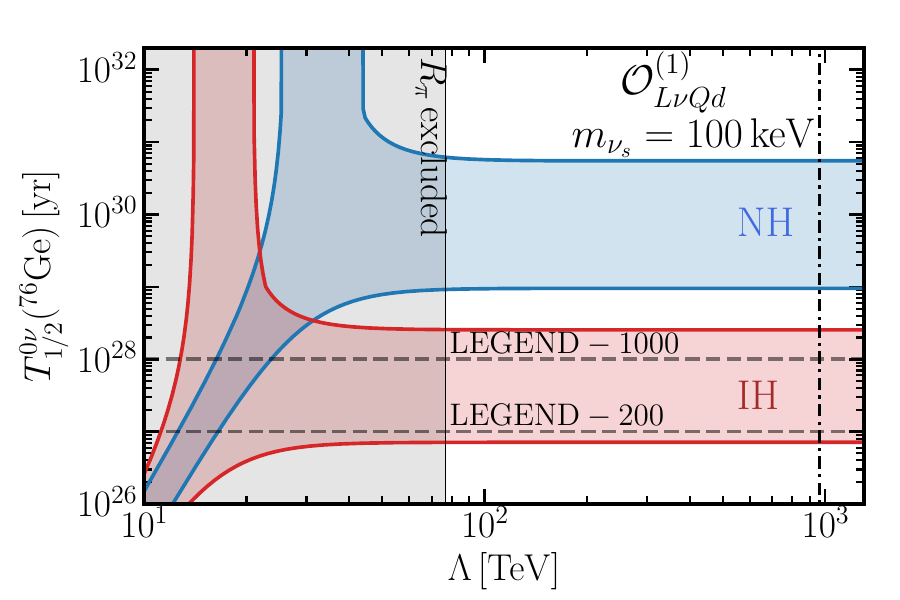}    
\end{tabular}
\vspace{-0.4cm}
\caption{
Half-life of neutrinoless double beta decay in $^{136}\mathrm{Xe}$ (left) and  $^{76}\mathrm{Ge}$ from the three light majorana neutrinos with $\mnu = 10^{-7}\,\mathrm{eV}$, and the ${\cal O}_{L\nu Qd}^{(1)}$ operator.
The bands include uncertainties arising from two different modelings of the NMEs, QRPA~\cite{Hyvarinen:2015bda} and Shell Model~\cite{Menendez:2017fdf}, as well as the variation of all PMNS angles and phases~\cite{Esteban:2020cvm}, and the dimension-$6$ phase.
For $\Lambda \gtrsim 40$ the $\tnu$ prediction is independent of the scalar operator ${\cal O}_{L\nu Qd}$. 
Horizontal dashed lines represent future experimental sensitivity and the grey shaded region shows the limit of Eq.~\eqref{eq:piondecay} from lepton universality of the pion decay.
The vertical dashed-dotted line shows the $X$-ray cancellation scale $\Lambda_e$ for NH from Tab.~\ref{tab:Xray_scale}.
}
\label{fig:DBD}
\end{figure}

We exemplarily study the ${\cal O}_{L\nu Qd}^{(1)}$ operator, but similar qualitative conclusions also hold for any operator leading to charged current interactions with first-generation quarks.
In the presence ${\cal O}_{L\nu Qd}^{(1)}$ we show the prediction of the neutrinoless double beta decay of Eq.~\eqref{eq:tnu} in Fig.~\ref{fig:DBD}.
For the plot we fix the sterile neutrino mass DM to be $100~\mathrm{keV}$. 
The bands shown in Fig.~\ref{fig:DBD}, blue for NH and red for IH, are obtained by varying the following input parameters:
i) The PMNS mixing angles and the Dirac CP phase within the $3\sigma$ region around their best-fit values obtained from global fits of neutrino oscillations data~\cite{Esteban:2020cvm};
ii) The two Majorana CP phases in the PMNS matrix are free to take any value between $0$ and $2 \pi$;
iii) The dimension-$6$ CP phase is varied between $0$ and $2 \pi$, unless the scale $\Lambda$ is around the $X$-ray cancellation scale of Tab.~\ref{tab:Xray_scale}, in which case the phase is aligned with the PMNS phases as shown in Eq.~\eqref{eq:cancellation_sol}; 
iv) We include uncertainties from calculations of the nuclear matrix elements (NMEs) and take into account two different models, the QRPA~\cite{Hyvarinen:2015bda} and the Shell Model~\cite{Menendez:2017fdf}.
We further neglect contribution to Eq.~\eqref{eq:tnu} proportional to the active-sterile mixing, which is certainly justified for the sterile neutrino DM candidate $\nu_s$ because its mixing needs to be smaller than the one given in Eq.~\eqref{eq:DW_contour} and thus its contribution to neutrinoless double beta decay is irrelevant.
On the other hand, the contribution from the two heavier sterile neutrinos, $\nu_4$ and $\nu_5$, can have some effect on $\tnu$, but this strongly depends on their mass and mixing to the active sector.
Since we are interested in the minimal scenario in which their mass is $> \mathcal{O}(\mathrm{GeV})$ and their mixing is on the seesaw line $U^2 \sim m_{\nu_a}/m_{\nu_{5/6}}$ the effect on $\tnu$ is negligible.
In summary, it is sufficient in our case to only sum over the three active neutrinos in Eq.~\eqref{eq:tnu}, i.e. $\sum_{i=1}^6 \mapsto \sum_{i=1}^3$.

We find that the contribution of the ${\cal O}_{L\nu Qd}$ operator to neutrinoless double beta decay, represented by $A_\nu^{\mathrm{d}=6}$ in Eq.~\eqref{eq:tnu}, dominates for $\Lambda \lesssim 30\,(15)~\mathrm{TeV}$ for NH (IH) over the standard dimension $4$ contribution.
This means that $\tnu$ in this region is much smaller than the half-life predicted if only dimension $4$ interactions were present.
For intermediate scales of $\Lambda \simeq 30-50 \,(15-30)\,\mathrm{TeV}$ for NH (IH), the contributions of the dimension $4$ and $6$ interactions are of the same order and the presence of CP-violating phases in both contributions can lead to cancellations on the amplitude level of Eq.~\eqref{eq:tnu}.
We note that this can happen for both NH and IH, contrary to the case of only dimension $4$ interactions where only in NH there can be a cancellation in the half-life.
For higher scales the dimension $6$ contribution decouples and the half-life is just given by the standard process of dimension $4$ Yukawa interactions.
The prediction of $\tnu$ in this regime thus becomes independent of $\Lambda$.
We further show experimental sensitivities in next-generation experiments, which can probe most of the IH region. The vertical dashed-dotted line corresponds to the scale required for the successful cancellation of X-ray contribution given in Tab.~\ref{tab:Xray_scale}.

Scalar operators can significantly affect pion decays, in particular the lepton flavor universality ratio    $ R_\pi = {\Gamma(\pi \rightarrow e \nu)}/{\Gamma(\pi \rightarrow \mu \nu)}$.
Neglecting the mass of the sterile neutrino, the operators  $C_{ Qu \nu L}$ and $C^{(1)}_{ L\nu Q d}$ induce the contributions
\begin{equation}
\frac{R_\pi}{R^{SM}_\pi} = \frac{1 + \frac{1}{V_{ud}^{2}} \left| \frac{m_\pi^2 v^2}{2 m_e (m_u + m_d)} \left(C_{\substack{ Qu \nu L \\ 114 e}} + C^{(1)*}_{\substack{L\nu Q d\\e411}}\right) \right|^2}{
1 + \frac{1}{V_{ud}^{2}} \left| \frac{m_\pi^2 v^2}{2 m_\mu (m_u + m_d)} \left(C_{\substack{ Qu \nu L \\ 114 \mu}} + C^{(1)*}_{\substack{L\nu Q d\\ \mu 411}}\right) \right|^2 }\,,
\end{equation}
where the operators are understood to be evaluated at the scale $\mu = 2\,\mathrm{GeV}$.  
The electron component, in particular, is enhanced by $1/m_e$, which leads to strong constraints.
Comparing with the experimental value
$    {R_\pi}/{R_\pi^{SM}} = 0.996 \pm 0.005$ \cite{Cirigliano:2013xha},
we get, for the electron component 
\begin{equation}\label{eq:piondecay}
   \left| C_{\substack{Q u \nu L\\ 114e}} (\mu = \Lambda)\right|, \left| C^{(1)}_{\substack{L \nu Q d\\ e411}} (\mu = \Lambda) \right| <  \frac{1}{\left(77 \, {\rm TeV}\right)^2}\,,
\end{equation}
under the assumption that there are no cancellations between both operators.
So in this sense Eq.~\eqref{eq:piondecay} represents the most optimistic scenario.
The constraint on the muonic components is weaker, $\Lambda \sim 6~\mathrm{TeV}$.
Both constraints agree with the results of Ref.~\cite{Fernandez-Martinez:2023phj}.
While the bound in Eq.~\eqref{eq:piondecay} is relatively strong, it is a factor of $10~(30)$ too weak to probe the cancellation regime in the NH (IH). 
The shaded region in Fig.~\ref{fig:DBD} is ruled out by this pion decay measurement and represents the highest scale we can currently probe the ${\cal O}_{L\nu Qd}^{(1)}$ operator.
The other charged current operators in Tab. \ref{tab:nuSMEFT_operators_dim6},
including the semileptonic tensor operator $C^{(3)}_{L \nu Q d}$, the right-handed $W$ coupling $C_{H\nu e}$ and the right-handed semileptonic operator $C_{du\nu e}$, are also constrained by $\beta$ decay at low energy \cite{Falkowski:2020pma,Dekens:2021qch} and by charged-current Drell-Yan at high-energy \cite{Cirigliano:2012ab}.
The BSM scale probed by these experiments is, however, 
close to 10 TeV \cite{Falkowski:2020pma,Dekens:2021qch,Cirigliano:2023nol}, too low to probe the cancellation regime.

The operators $\mathcal O_{L\nu Le}$, 
$\mathcal O^{(1,3)}_{L\nu Q d}$ can generate contributions to the electron and nucleon EDM. At one loop, these operators induce the electron and down dipole operators
\begin{align}
    \dot C_{\substack{e W\\ pr}} &= - \frac{g_2}{4} C_{\substack{L \nu L e \\ p s t r}} \left[Y_\nu^\dagger\right]_{s t}\,, &\qquad 
        \dot C_{\substack{e B\\ p r }} &= \frac{g_1 {\rm y}_{\ell}}{2}  C_{\substack{L \nu L e \\ p s t s}} \left[Y^\dagger_\nu\right]_{ts}\,,\\
        \dot C_{\substack{d W\\ pr}} &=  2  g_2 C^{(3)}_{\substack{L \nu Q q \\ p s t r}} \left[Y_\nu^\dagger\right]_{s t}\,, &\qquad 
        \dot C_{\substack{d B\\ p r }} &= - 4 g_1 {\rm y}_{\ell}  C^{(3)}_{\substack{L \nu Q d \\ p s t s}} \left[Y^\dagger_\nu\right]_{ts}\,.
\end{align}
While these combinations mainly generate $Z$ dipoles, the LL evolution of the $C_{e W}$, $C_{e B}$ and $C_{d W}$, $C_{d B}$ also results in photon dipoles. 
Since they are proportional to the neutrino Yukawa, however, these contributions are very small
\begin{align}
    d_e &= - 2 \frac{v}{\sqrt{2} \Lambda^2} \textrm{Im}  (s_w C_{\substack{eW\\ ee}} + c_w C_{\substack{eB\\ ee}}) \approx   \frac{e m_{\nu_s}}{(4\pi \Lambda)^2} \frac{e^2}{(4\pi)^2} \log^2 \frac{m_Z}{\Lambda}  C_{\substack{ L\nu Le\\ e 4 e e}}  U^*_{e 4}   \\
    & \approx 2 \cdot 10^{-30}  \frac{m_{\nu_s}}{100\, {\rm keV}} \left(\frac{10 \, {\rm TeV}}{\Lambda}  \right)^2 U^*_{e 4} \, \, e\,{\rm cm}\,.
\end{align}
With the current limit, $|d_e| < 4.0 \cdot 10^{-30}$ $e$ cm~\cite{Roussy:2022cmp}, electron EDM searches would only be sensitive to $U_{e 4}\sim \mathcal O(1)$, for $m_{\nu_s} = 100\,\mathrm{keV}$ and BSM scales of $\Lambda = 10\,\mathrm{TeV}$. For $\Lambda \sim 10^3\,\mathrm{TeV}$ and $U_{e4} \sim 10^{-6}$, as in Tab.~\ref{tab:Xray_scale}, the electron EDM would be of order $10^{-40}$ $e$ cm, swamped by the SM background.

In conclusion, contributions from dimension $6$ operators that succeed in a cancellation of the $X$-ray emission as well as an explanation of DM abundance in form of sterile neutrinos, are hidden in $0\nu\beta\beta$ process. Therefore, in order to probe the cancellation scenario, we need to investigate other probes, including beta decays and multi-species EDM searches, that can give distinctive predictions between dimension $4$ and $6$ contributions. Any discovery in next-generation experiments that would point to lower scale than those in Tab.~\ref{tab:Xray_scale} and~\ref{tab:Xray_scale_no_mixing} could falsify the cancellation mechanism.


\section{Conclusions}
\label{sec:conclu}
We have studied the generation of $\mathcal{O}(\mathrm{keV})$ sterile neutrino dark matter (DM) within the framework of the neutrino-extended Standard Model effective field theory ($\nu$SMEFT).
We include dimension $5$ and $6$ interactions and show that they can generically lead to a relic abundance of sterile neutrinos that matches the observed DM density.
Because the $\nu$SMEFT includes tree-level effective interactions between sterile neutrinos and SM particles, the production mechanism is effective even for negligible active-sterile neutrino mixing, unlike in the Dodelson-Widrow mechanism~\cite{Dodelson:1993je} and many of its extensions~\cite{DeGouvea:2019wpf, Kelly:2020pcy, Fuller:2024noz, Johns:2019cwc, Astros:2023xhe, Bringmann:2022aim, An:2023mkf, Koutroulis:2023fgp, Holst:2023hff, Jaramillo:2022mos, Benso:2021hhh, Cho:2021yxk, Shi:1998km, Benso:2019jog}.
On the other hand, special attention has been paid to the large mixing regime in which the traditional dimension $4$ setup faces strong constraints arising from $X$-ray limits on the radiative decay $\nu_s \to \nu_a \gamma$.
We use the full $1$-loop renormalization group running and $1$-loop matching of $\nu$SMEFT on its low energy companion, $\nu$LEFT, at the electroweak (EW) scale and identify the set of $\nu$SMEFT operators which create a neutrino photon dipole.
We find that seven out of the sixteen dimension-$6$ $\nu$SMEFT operators in Tab.~\ref{tab:nuSMEFT_operators_dim6} can induce sizable neutrino photon dipole operators, and can lead to sufficient destructive interference with the dimension-$4$ contribution to evade current $X$-rays limits for values of the new physics scale $\Lambda$ well beyond the EW scale.
In particular, the right-handed charged current operator $\mathcal O_{H\nu e}$, the dipole operators $\mathcal O_{\nu W}$ and $\mathcal O_{\nu B}$,
the purely leptonic scalar operator $\mathcal O_{L\nu Le}$, the scalar and tensor semileptonic operators $\mathcal O^{(1,3)}_{L\nu Qd}$, and the vector semileptonic operator $\mathcal O_{d u \nu e}$ induce significant neutrino dipole operators.

By solving the renormalization group equations, we find unique values of the $\nu$SMEFT operators scale for a given pair of $(\mnu,\mnus)$, while the $d>4$ CP-violating phases are aligned with the PMNS phases in the region of interest (see Tab.~\ref{tab:Xray_scale}). 
This opens the parameter space of active-sterile neutrino mixings of up to $4$ orders of magnitude larger than naively expected (see Fig.~\ref{fig:cons_cosmo}).
The values of $\Lambda$ required by the cancellation are typically very large, from a minimum of $750\,\mathrm{TeV}$ to a maximum of $3\cdot 10^4\,\mathrm{TeV}$, for $m_{\nu_{\rm lightest}} = 10^{-7}\,\mathrm{eV}$. 
Interestingly, to match the observed DM density we predict a low-scale re-heating temperature of the order $\mathcal{O}(10^2\,\mathrm{GeV})$ (see Tab.~\ref{tab:Xray_scale}). We also find that the DM density is successfully explained by $\nu$SMEFT operators even in the zero-mixing limit. In this limit, the BSM scale and, consequently, the re-heating temperature, of dipole-inducing operators needs to be 10 times higher (see Tab.~\ref{tab:Xray_scale_no_mixing}).
The considerably larger parameter space can also allow for sterile neutrino DM in the $\mathcal{O}(\mathrm{MeV})$ range (or higher), which will be studied elsewhere.

We finally discuss laboratory probes of relevant $\nu$SMEFT operators, such as the precision measurement of 
lepton flavor universality ratios in
pion decay, and searches for $0\nu\beta\beta$ and EDMs. 
These experiments can reach, at best, $\Lambda \lesssim 100\,\mathrm{TeV}$, far beneath the scales required  
to avoid $X$-ray bounds. 
We thus find that, in sterile neutrino DM scenarios, the 
parameter space for 
the dimension-6 operators that induce large neutrino photon dipoles is very constrained 
and leaves no room for observation in upcoming terrestrial experiments.
On the other hand, evidence for $\mathrm{keV}$ sterile neutrinos in the next generation of  experiments, with non-standard interactions of the type discussed above, will immediately rule out the $\nu$SMEFT sterile neutrino DM scenario.


\begin{acknowledgments}

We thank G. Fuller, L. Johns and J. L\'opez-Pav\'on for useful discussions and/or clarifications.
This work was supported by the US Department of Energy Office 
and by the Laboratory Directed Research and Development (LDRD) program of Los Alamos National Laboratory under project numbers
20230047DR, 20230408ER, 20220706PRD1.
Los Alamos National Laboratory is operated by Triad
National Security, LLC, for the National Nuclear Security Administration of the U.S. Department of Energy
(Contract No. 89233218CNA000001).
The authors gratefully acknowledge the computer resources at Artemisa and the technical support provided by the Instituto de Fisica Corpuscular, IFIC (CSIC-UV). 
Artemisa is co-funded by the European Union through the 2014-2020 ERDF Operative Programme of Comunitat Valenciana, project IDIFEDER/2018/048.

\end{acknowledgments}


\appendix


\section{Appendix: Active neutrino Yukawa and active-sterile mixing}
\label{app:casas_ibarra}
This appendix is devoted to a short derivation of the general parametrization of the neutrino Yukawa couplings as well as the active to sterile mixing matrix, which directly accounts for the constraints arising from neutrino oscillations experiments.

The complete neutrino mass matrix $M_\nu$, see Eq.~\eqref{eq:Mnu_def} can be diagonalized by a unitary matrix $U$ via
\begin{align}
    M_\nu = U^* \mathrm{diag}(m,M) U^\dagger\,.
\end{align}
Consider the limit $m_L \to 0$ in $M_\nu$, which is the relevant limit for this work, as shown in Sec.~\ref{subsubsec:xray_cancellation}.
We can then re-organize Eq.~\eqref{eq:mL_diagonalization} as
\begin{align}
    \mathbb{1}  &= - m^{-1/2} \pmns^\dagger \theta M^{1/2} M^{1/2} \theta^T \pmns^* m^{-1/2} \\
                &= - \left[M^{1/2} \theta^T \pmns^* m^{-1/2}  \right]^T \left[ M^{1/2} \theta^T \pmns^* m^{-1/2} \right]\,, 
\end{align}
which defines the complex, orthogonal matrix 
\begin{align}
\label{app:eq:R_def}
    R^* = - i M^{1/2} \theta^T \pmns^* m^{-1/2}\,.
\end{align}
This matrix has a shape of $n\times 3$ with $n$ being the number of sterile neutrino species. 
If $n=3$, which is the scenario considered in the main text, the matrix generically can be parametrized as a rotation matrix depending on three complex angles $z_{ij}$
\begin{eqnarray}
\label{app:eq:r_matrix}
R = \left( \begin{array}{ccc}
\cos z_{12} & -\sin z_{12} & 0 \\
\sin z_{12} & \cos z_{12} & 0 \\
0& 0 &1\end{array}\right)
\left( \begin{array}{ccc}
\cos z_{13} & 0& -\sin z_{13}\\ 
0 & 1 & 0 \\
\sin z_{13} & 0 &\cos z_{13}\\
 \end{array}\right) 
\left( \begin{array}{ccc}
1 & 0 & 0 \\
0& \cos z_{23} & -\sin z_{23} \\
0& \sin z_{23}  &\cos z_{23}\\
\end{array}\right).
\end{eqnarray}
Thus, by solving Eq.~\eqref{app:eq:R_def}, the light to sterile mixing $\theta$ can be parametrized via
\begin{align}
\label{app:eq:Mixing_casas_ibarra}
    \theta = i \pmns \sqrt{m} R^\dagger M^{-1/2}\,.
\end{align}
\noindent
Using Eq.~\eqref{eq:mD_diagonalization}, together with the fact that $m_D = Y_\nu\, v/\sqrt{2} $, leads to
\begin{align}
\label{app:eq:Yukawa_casas_ibarra}
    Y_\nu = i \frac{\sqrt{2}}{v} \pmns^{\phantom{T}} \sqrt{m}  R^\dagger \sqrt{M}\,.
\end{align}


\section{Appendix: Matrix elements}
\label{app:M2}
This appendix collects the matrix elements for the $\nu$SMEFT operators considered in Sec.~\ref{subsubsec:xray_cancellation}, representing the crucial particle physics input to determine the sterile neutrino number density evolution via the Boltzmann equation~\eqref{eq:boltzmann_eq_nus}.
Following the definition of Eq.~\eqref{eq:boltzmann_eq_nus}, the matrix elements summarized here are summed over initial and final state spins, but not averaged.
When quarks are considered, the appropriate color factor $N_c= 3$ has to be included.
The light neutrino masses $\mnua$ are considered to be zero.

\begin{figure}[!t]
\centering
\begin{tabular}{c}
\hspace{-0.3cm} \includegraphics[width=0.96\textwidth]{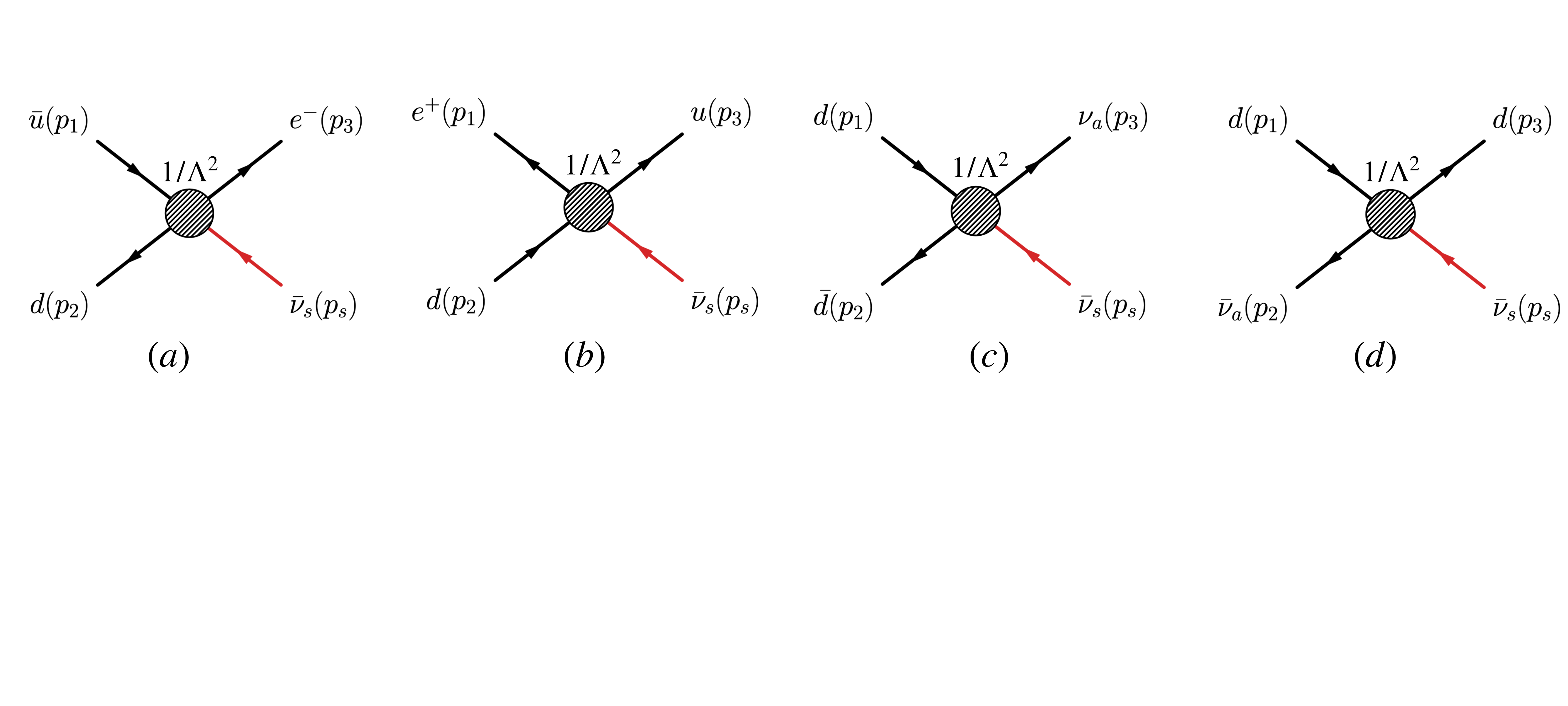} 
\end{tabular}
\vspace{-3.0cm}
\caption{
Feynman diagrams for the $\nu$SMEFT operator $\bar{L}^i \nu_R \epsilon_{ij} \bar{Q}^j d_R$ and $\bar{L}^i \sigma^{\mu\nu} \nu_R \epsilon_{ij} \bar{Q}^j \sigma_{\mu\nu} d_R$ producing sterile antineutrinos $\bar{\nu}_s$.
The operator $\bar{d}_R \gamma^{\mu} u_R \bar{v}_R \gamma_{\mu} e_R$ realizes only the charged current topologies $(b)$ and $(b)$.
}
\label{app:fig:feynman_LnuQd}
\end{figure}

The minimal set of Feynman diagrams leading to a sterile neutrino production are shown in Fig.~\ref{fig:feynman_lnule} for the operator $\bar{L}^i \nu_R \epsilon_{ij} \bar{L}^j e_R$.
In Fig.~\ref{app:fig:feynman_LnuQd} the Feynman diagrams are shown for the operators $\bar{L}^i \nu_R \epsilon_{ij} \bar{Q}^j d_R$, $\bar{L}^i \sigma^{\mu\nu} \nu_R \epsilon_{ij} \bar{Q}^j \sigma_{\mu\nu} d_R$ and $\bar{d}_R \gamma^{\mu} u_R \bar{v}_R \gamma_{\mu} e_R$.

In the following the matrix elements are expressed in terms of the Mandelstam variables $s, t$, associated to each topology as given by the corresponding Feynman diagram.
We can further express $t$ as a function of $s$ and $\theta$, with $\theta$ being the center-of-mass angle between particle $1$ and $3$~\cite{ParticleDataGroup:2022pth}
\begin{align}
    t = t_0 - 4 \sin ^2\left(\frac{\vartheta }{2}\right) \sqrt{\frac{\left(m_1^2-m_2^2+s\right)^2}{4 s}-m_1^2} \sqrt{\frac{\left(m_3^2-m_4^2+s\right)^2}{4 s}-m_3^2}\,,
\end{align}
with 
{\small 
\begin{align}
    t_0 = \frac{\left(m_1^2-m_2^2-m_3^2+m_4^2\right)^2}{4 s}-\left(\sqrt{\frac{\left(m_1^2-m_2^2+s\right)^2}{4 s}-m_1^2}-\sqrt{\frac{\left(m_3^2-m_4^2+s\right)^2}{4 s}-m_3^2}\right)^2\,.
\end{align}
}
Thus all found $|\mathcal{M}|^2$ can be directly integrated over $\mathrm{d}\Omega$.

\noindent
\textbf{$\mathbf{\bar{L}^i \nu_R \epsilon_{ij} \bar{L}^j e_R}$.}\,
For topology $(a)$ and $(b)$ 
\begin{align}
    |\mathcal{M}|^2 = \frac{\mnus^2 \left(5 m_e^2-s-t\right)+m_e^4-2 m_e^2 (s+t)+(s-t)^2}{\Lambda ^4}\,.
\end{align}

\noindent
\textbf{$\mathbf{\bar{L}^i \nu_R \epsilon_{ij} \bar{Q}^j d_R}$.}\,
For topology $(a)$ 
\begin{align}
    |\mathcal{M}|^2 = N_c \frac{\left(\mnus^2+m_e^2-s\right) \left(m_d^2+m_u^2-s\right)}{\Lambda ^4} \,.
\end{align}
For topology $(b)$ 
\begin{align}
    |\mathcal{M}|^2 = N_c \frac{\left(m_e^2+\mnus^2-s-t\right) \left(m_d^2+m_u^2-s-t\right)}{\Lambda ^4}
\end{align}
For topology $(c)$ 
\begin{align}
    |\mathcal{M}|^2 = N_c \frac{\left(2 m_d^2-s\right) \left(\mnus^2-s\right)}{\Lambda ^4}
\end{align}
For topology $(d)$ 
\begin{align}
    |\mathcal{M}|^2 = N_c\frac{\left(2 m_d^2-t\right) \left(\mnus^2-t\right)}{\Lambda ^4}
\end{align}

\noindent
\textbf{$\mathbf{\bar{L}^i \sigma^{\mu\nu} \nu_R \epsilon_{ij} \bar{Q}^j \sigma_{\mu\nu} d_R}$.}\,
For topology $(a)$ 
{\normalsize
\begin{align}
\begin{split}
    |\mathcal{M}|^2 &= N_c \frac{16}{\Lambda^4} \left(m_u^2 \left(4 m_d^2+m_e^2+\mnus^2-s-4 t\right)+m_d^2 \left(m_e^2+\mnus^2-s-4 t\right) \right. \\
                    &\left.-4 t \left(m_e^2+\mnus^2-s\right)+4 m_e^2 \mnus^2-m_e^2 s-\mnus^2 s+s^2+4 t^2\right) \,.
\end{split}
\end{align}
}
For topology $(b)$ 
\begin{align}
\begin{split}
    |\mathcal{M}|^2 &= N_c \frac{16}{\Lambda^4}\left(m_e^2 \left(m_d^2+m_u^2+4 \mnus^2-s-t\right)+m_d^2 \left(4 m_u^2+\mnus^2-s-t\right) \right. \\
                    &\left. -t \left(m_u^2+\mnus^2+2 s\right)+\left(m_u^2-s\right) \left(\mnus^2-s\right)+t^2\right) \,.
\end{split}
\end{align}
For topology $(c)$ 
\begin{align}
    |\mathcal{M}|^2 = N_c \frac{16 \left(4 m_d^4+2 m_d^2 \left(\mnus^2-s-4 t\right)-\mnus^2 (s+4 t)+(s+2 t)^2\right)}{\Lambda ^4} \,.
\end{align}
For topology $(d)$ 
\begin{align}
    |\mathcal{M}|^2 = N_c \frac{16 \left(4 m_d^4+2 m_d^2 \left(\mnus^2-4 s-t\right)-\mnus^2 (4 s+t)+(2 s+t)^2\right)}{\Lambda ^4} \,.
\end{align}

\noindent
\textbf{$\mathbf{\bar{d}_R \gamma^{\mu} u_R \bar{v}_R \gamma_{\mu} e_R}$.}\,
For topology $(a)$ and $(b)$ 
\begin{align}
    |\mathcal{M}|^2 = N_c \frac{4 \left(m_1^2+m_3^2-t\right) \left(m_2^2+\mnus^2-t\right)}{\Lambda ^4}\,,
\end{align}
with $m_1= m_u, m_2= m_d, m_3= m_e$ for topology $(a)$ and for $(b)$ $m_1= m_e, m_2= m_d, m_3= m_u$.


\section{Appendix: Non perturbative dipole contributions}
\label{app:np}
In this Appendix, we briefly discuss the nonperturbative contributions to $\nu_s \rightarrow \nu_a \gamma$ induced by tensor operators.
By using the results of Ref.~\cite{Cirigliano:2021img}, this contribution can be captured by shifting the coefficient of the dipole operator as
\begin{align}
\label{app:eq:np}
    C_{\substack{\nu F \\r s}}( \mu_{\rm low}) \rightarrow -2 e Q_d \frac{v^2}{\Lambda^2} \left(
    C^{T, RR}_{\substack{\nu d \\ rs11}}(\mu_{\rm low}) + C^{T, RR}_{\substack{\nu d \\ rs22}}(\mu_{\rm low}) \right)\frac{i \Pi_{VT}(0)}{v}\,,
\end{align}
where $\Pi_{VT}(0)$ is the correlator of the vector and tensor currents at zero momentum (see Eq. (D47) in Ref.~\cite{Cirigliano:2021img}  for the precise definition).
Adopting a large-$N_C$ inspired  resonance model~\cite{Cata:2008zc,Cirigliano:2021img}, $\Pi_{VT}(0)$ is given by
\begin{equation}
    \frac{\Pi_{VT}(0)}{v} = \frac{B_0 F_\pi^2}{v M_V^2} \sim 1.6 \cdot  10^{-4}, 
\end{equation}
where $B_0 = 2.8$ GeV at the scale $\mu_{\rm low} = 2\,\mathrm{GeV}$, 
$F_\pi = 92.2$ MeV and $M_V = 770$ MeV.
Eq. \eqref{app:eq:np} should be compared with the perturbative contribution
\begin{equation}
     C_{\substack{\nu F \\ rs}}(\mu_{\rm low}) \sim -\frac{e Q_d 8 N_C}{16 \pi^2}  \frac{v^2}{\Lambda^2}\left(\frac{m_d}{v} C^{T, RR}_{\substack{\nu d \\ rs11}} + \frac{m_s}{v} C^{T, RR}_{\substack{\nu d \\ rs22}} \right) \log \frac{\mu_{\rm low}}{\mu_{\rm high}}. 
\end{equation}
For the $d$ quark, the nonperturbative contribution 
is  larger by approximately a factor of 10. For the $s$ quark, the perturbative contribution is larger.


\newpage
\bibliography{biblio}

\end{document}